\documentclass[12pt, reqno]{amsart}
\usepackage{amsmath,amssymb,pdflscape,bm,xr}
\usepackage{amsfonts,natbib}
\usepackage{graphicx,epsfig,setspace}
\usepackage[margin=1in]{geometry}

\numberwithin{equation}{section}
\begin{document}

\newcommand{\hmmv}{\hat{\mathbb{V}}}
\newcommand{\neyman}{\mathrm{Neyman}}
\newcommand{\aronow}{\mathrm{AGL}}
\newcommand{\sboot}{\mathrm{s-boot}}
\newcommand{\cboot}{\mathrm{c-boot}}
\newcommand{\nsboot}{\mathrm{neyman-s-boot}}
\newcommand{\ncboot}{\mathrm{neyman-c-boot}}
\newcommand{\acboot}{\mathrm{aronow-c-boot}}
\newcommand{\tock}{?}
\newcommand{\oy}{\overline{Y}}
\newcommand{\pr}{\mathrm{pr}}
\newcommand{\tick}{\checkmark}
\newcommand{\br}{\mathbf{R}}
\newcommand{\brr}{\mathbf{r}}
\newcommand{\bw}{\mathbf{W}}
\newcommand{\bww}{\mathbf{w}}
\newcommand{\ct}{0}
\newcommand{\tc}{1}
\newcommand{\cov}{\textnormal{Cov}}
\newcommand{\var}{\textnormal{Var}}
\newcommand{\diag}{\textnormal{diag}}
\newcommand{\plim}{\textnormal{plim}_n}
\newcommand{\dum}{1\hspace{-2.5pt}\textnormal{l}}
\newcommand{\ind}{\bot\hspace{-6pt}\bot}
\newcommand{\co}{\textnormal{co}}
\newcommand{\tr}{\textnormal{tr }}
\newcommand{\fsgn}{\textnormal{\footnotesize sgn}}
\newcommand{\sgn}{\textnormal{sgn}}
\newcommand{\fatb}{\mathbf{b}}
\newcommand{\fatp}{\mathbf{p}}
\newcommand{\trace}{\textnormal{trace}}
\newcommand{\mmv}{\mathbb{V}}

\newtheorem{dfn}{Definition}[section]
\newtheorem{rem}{Remark}[section]
\newtheorem{cor}{Corollary}[section]
\newtheorem{thm}{Theorem}[section]
\newtheorem{lem}{Lemma}[section]
\newtheorem{notn}{Notation}[section]
\newtheorem{con}{Condition}[section]
\newtheorem{prp}{Proposition}[section]
\newtheorem{pty}{Property}[section]
\newtheorem{ass}{Assumption}[section]
\newtheorem{ex}{Example}[section]
\newtheorem*{cst1}{Constraint S}
\newtheorem*{cst2}{Constraint U}
\newtheorem{qn}{Question}[section]

%\singlespacing

\onehalfspacing
\title[Causal Bootstrap]{A Causal Bootstrap}
\author[Guido Imbens and Konrad Menzel]{Guido Imbens$^{\sharp}$ \hspace{0.6cm}Konrad Menzel$^{\dag}$}
%\author[1]{Guido Imbens}\affil[1]{Stanford GSB}\author[2]{Konrad Menzel}\affil[2]{New York University}
%\renewcommand\Authands{ and }
\date{March 2018 - this version December 2018.}\thanks{$^{\sharp}$Guido Imbens: Stanford GSB, Email: Imbens@stanford.edu, $^{\dag}$Konrad Menzel: New York University, Email: km125@nyu.edu. We thank Alberto Abadie and Alfred Galichon for useful conversations.}

\begin{abstract}
The bootstrap, introduced by \citet{Efr82}, has become a very popular method for estimating variances and constructing confidence intervals. A key insight is that one can approximate the properties of estimators by using the empirical distribution function of the sample as an approximation for the true distribution function. This approach views the uncertainty in the estimator as coming exclusively from sampling uncertainty. We argue that for causal estimands the uncertainty arises entirely, or partially, from a different source, corresponding to the stochastic nature of the treatment received. We develop a bootstrap procedure that accounts for this uncertainty, and compare its properties to that of the classical bootstrap.
\\[4pt]
\end{abstract}

\maketitle

% Structure:

\section{Introduction}

\label{sec:introduction}

%\subsection{Problem Description}

The bootstrap, introduced by \citet{Efr82}, has become a very popular method for constructing hypothesis tests or confidence intervals. This popularity stems in part from the fact that it provides approximations to the distribution of an estimator or statistic that are in certain cases superior to those obtained from using a Gaussian asymptotic approximation together with estimated standard errors (asymptotic refinement). While the classical bootstrap is designed to approximate distributions that result from repeated sampling from a large population, this paper shows how to adapt the bootstrap principle when the estimand of interest is a causal parameter, and the data is generated by a randomized experiment.

Permutation tests, such as Fisher's exact test (see e.g. \cite{IRu15}), can yield exact p-values under the auxiliary hypothesis that treatment effects are constant across units, however we argue below that those methods are not suitable for forming confidence intervals for parameters describing the distribution of causal effects in a given population. For the average treatment effect, causal standard errors have been proposed by \cite{AGL14} as well as \cite{AAIW17}. These methods impose no restrictions on treatment effect heterogeneity but their use generally relies on a Gaussian limiting approximation. We propose a bootstrap approach to causal inference which also does not restrict treatment heterogeneity, but improves on the Gaussian asymptotic approximation in samples of small or moderate size.

%may yield asymptotic refinements to

%Causal standard errors, which have been derived for this type of problem in the previous literature - see e.g. \cite{Rob88}, \cite{AGL14}, and our discussion below - can be used for these inference tasks in combination with a Gaussian limiting approximation to the distribution of an estimator for that parameter.

Using the potential outcome framework, e.g., \cite{IRu15}, we are interested in the average causal effect of a binary variable $W_i\in\{0,1\}$ (the ``treatment") on an outcome variable whose potential outcomes we denote with $Y_i(0),Y_i(1)$, for a population of $N$ units $i=1,\dots,N$. Implicitly we assume that the potential outcomes $Y_i(w)$ for unit $i$ do not vary with the treatment status assigned to other units, known as the Stable Unit Treatment Value Assumption (SUTVA, \citet{Rub78}). For all units in the population we observe the treatment $W_i$ and the realized outcome $Y_i:=Y_i(W_i)$. One common estimand is the average effect for the $N$ units in the population:
\begin{equation}\label{tau_defn} \tau_{ATE}:=\frac{1}{N}\sum_{i=1}^N \Bigl(Y_i(1)-Y_i(0)\Bigr).\end{equation}
We assume that the data arise from a completely randomized experiment, where $n\leq N$ units are selected at random from the population as experimental subjects, of which $n_1$ units are then randomly assigned to receive the treatment, and the remaining $n_0 = n - n_1$ units are assigned to the control group. We let $R_i\in\{0,1\}$ denote an indicator whether the $i$th unit is included in the sample.

Specifically, for $\br :=(r_1,\dots,r_N)'$ and $\bw:=(w_1,\dots,w_N)'$ we have
\begin{eqnarray}
\nonumber\pr(\br=\brr)&=&\left\{\begin{array}{ccl}\binom{N}{n}^{-1}&\hspace{0.3cm}&\textnormal{if }\brr\in\{0,1\}^N\textnormal{ and }\sum r_i=n\\0&&\textnormal{otherwise}\end{array}\right.\\
\nonumber\pr(\bw=\bww|\br=\brr)&=&\left\{\begin{array}{ccl}\binom{n}{n_1}^{-1}&\hspace{0.3cm}&\textnormal{if }\sum r_iw_i=n_1,\;\bww\in\{0,1\}^N\textnormal{ and }(1-r_i)w_i=0\textnormal{ for all }i\\
0&&\textnormal{otherwise}\end{array}\right.
\end{eqnarray}

\subsection{Sampling Uncertainty and Design Uncertainty}

We wish to distinguish between two types of uncertainty in estimators, \emph{sampling uncertainty} arising from the stochastic nature of $\mathbf{R}$, and \emph{design uncertainty} arising from the stochastic nature of $\mathbf{W}$.

To characterize sampling uncertainty we postulate the existence of a large, possibly infinite, population. We draw a random sample from this population, and observe for each unit in this sample a set of values, say, a pair $(Y_i,W_i)$. We may be interested in the difference between the population averages of $Y_i$ for the subpopulations with $W_i=0$ and $W_i=1$. We can estimate this object using the difference in average outcomes by $W_i$ values in the sample. This estimator differs from the target because we do not observe all units in the population. Had we drawn a different random sample, with different units, the value of the estimator would have been different. See Table \ref{tabel_sampling}, where $R_i$ is the sampling indicator, equal to 1 for sampled units and 0 otherwise. The uncertainty arising from the randomness in $\mathbf{R}$ is captured by the conventional standard error.

\begin{table}[ht]
\caption{\textsc{: Sampling-based Uncertainty  ($\tick$ is observed, $\tock$ is missing)}}
\label{tabel_sampling}\vskip1cm
\begin{center}
\begin{tabular}{c|cccccccccccccc}
  & \multicolumn{3}{c}{Actual} &&
 \multicolumn{3}{c}{Alternative}&&
 \multicolumn{3}{c}{Alternative}&& $\hdots$\\
Unit & \multicolumn{3}{c}{Sample} &&
 \multicolumn{3}{c}{Sample I}&&
 \multicolumn{3}{c}{Sample II}&& $\hdots$\\
 & $Y_i$ & $W_i$&$R_i$&\hskip0.5cm \ \ \ & $Y_i$ & $W_i$ & $R_i$
&\hskip0.5cm \ \ \ & $Y_i$ & $W_i$ & $R_i$&\hskip0.5cm& $\hdots$
\\
 \hline \\
$1$ & $\tick$ & $\tick$&1& & $\tock$ & $\tock$&0& & $\tock$ & $\tock$&0&& $\hdots$\\
$2$ & $\tock$ & $\tock$&0& & $\tock$ & $\tock$&0& & $\tock$ & $\tock$&0&& $\hdots$\\
$3$ & $\tock$ & $\tock$&0& & $\tick$ & $\tick$&1& & $\tick$ & $\tick$&1&& $\hdots$\\
$4$ & $\tock$ & $\tock$&0&& $\tick$ & $\tick$&1& & $\tock$ & $\tock$&0&& $\hdots$\\
\vdots & \vdots &\vdots& \vdots& & \vdots & \vdots&\vdots& & \vdots & \vdots&\vdots&& $\hdots$\\
$N$ & $\tick$ & $\tick$&1&& $\tock$ & $\tock$&0& & $\tock$ & $\tock$&0&& $\hdots$
\end{tabular}
\end{center}
\end{table}

In a randomized experiment the uncertainty is not necessarily of this sampling variety. Instead we can think of the uncertainty arising from the stochastic nature of the assignment, $\mathbf{W}$. For units with $W_i=0$ we observe $Y_i(0)$, and for units with $W_i=1$ we observe $Y_i(1)$. In our sample units have a particular set of assignments. In a repeated sampling thought experiment the units in the sample would have remained the same, but their assignments would might been different, leading to a different value for the estimator. See Table \ref{tabel_assignment}.

\begin{table}[ht]
\caption{\textsc{: Design-based Uncertainty ($\tick$ is observed, $\tock$ is missing)}}
\label{tabel_assignment}\vskip1cm
\begin{center}
\begin{tabular}{c|cccccccccccccc}
  & \multicolumn{3}{c}{Actual} &&
 \multicolumn{3}{c}{Alternative}&&
 \multicolumn{3}{c}{Alternative}&& $\hdots$\\
Unit & \multicolumn{3}{c}{Sample} &&
 \multicolumn{3}{c}{Sample I}&&
 \multicolumn{3}{c}{Sample II}&& $\hdots$\\
& $Y_i(1)$ & $Y_i(0)$&  $W_i$ &\hskip0.5cm \ \ \  & $Y_i(1)$ & $Y_i(0)$& $W_i$ &\hskip0.5cm \ \ \  & $Y_i(1)$ & $Y_i(0)$& $W_i$ &\hskip0.5cm& $\hdots$\\
 \hline \\
$1$ & $\tick$ & $\tock$&1& & $\tick$ & $\tock$&1& & \tock & \tick&0&& $\hdots$\\
$2$ & $\tock$ & $\tick$&0& & $\tock$ & $\tick$&0& & \tock & \tick&0&& $\hdots$\\
$3$ & $\tock$ & $\tick$&0& & $\tick$ & $\tock$&1& & \tick& \tock&1&& $\hdots$\\
$4$ & $\tock$ & $\tick$&0& & $\tock$ & $\tick$&0& & \tick & \tock&1&& $\hdots$\\
%5 & $\tick$ & $\tock$& & $\tock$ & $\tick$& & \tick & \tock&& $\hdots$\\
\vdots & \vdots & \vdots&\vdots && \vdots&\vdots & \vdots && \vdots&\vdots&\vdots&& $\hdots$\\
$N$ & $\tick$ & $\tock$&1& & $\tock$ & $\tick$&0& & \tock & \tick&0&& $\hdots$
\end{tabular}
\end{center}
\end{table}

\subsection{Notation} In the following, we denote the distribution of potential outcomes in the population with $F_{01}^p(y_0,y_1):=\frac1N\sum_{i=1}^N\dum\{Y_i(0)\leq y_0,Y_i(1)\leq y_1\}$ and the size of that population with $N$. The distribution in the sample of size $n$ is denoted with $F_{01}^s(y_0,y_1):=\frac1n\sum_{i=1}^NR_i\dum\{Y_i(0)\leq y_0,Y_i(1)\leq y_1\}$. $F_{01}^s(\cdot,\cdot)$ is not quite the empirical distribution because we only observe one of the values in each pair $(Y_i(0),Y_i(1))$. We use $p$ superscripts throughout to indicate population quantities, and $s$ superscripts to denote their sample analogs. The number of treated units in the sample is denoted with $n_1$, the number of control units with $n_0=n-n_1$, and the respective shares of treated and control units with $p:=n_1/n$, so that $1-p=n_0/n$. We also define the empirical c.d.f. for either potential outcome given the randomized treatment as $\hat{F}_{0}(y_0):=\frac1{n_0}\sum_{i=1}^NR_i(1-W_i)\dum\{Y_i(0)\leq y_0\}$ and $\hat{F}_{1}(y_1):=\frac1{n_1}\sum_{i=1}^NR_iW_i\dum\{Y_i(1)\leq y_1\}$.

\section{The Causal Bootstrap for Average Treatment Effects}
\label{sec:ate}

In this section we consider causal bootstrap inference for the population average treatment effect $\tau_{ATE}$ defined in (\ref{tau_defn}). The estimator we use is the difference in sample averages by treatment status:
\[ \hat\tau_{ATE}:=\oy_\tc-\oy_\ct,\]
where
\[\oy_\tc:=\frac{1}{n_1}\sum_{i=1}^N R_iW_i Y_i,\hskip1cm {\rm and}\ \
\oy_\ct:=
\frac{1}{n_0}
\sum_{i=1}^N R_i(1-W_i) Y_i.
\]
The repeated sampling perspective we take is one where the potential outcomes $(Y_i(0),Y_i(1))$ are fixed for all $N$ units in the population. The stochastic properties of the estimator arise from the stochastic nature of the assignment and sampling, which are both sources of randomness in the average of realized outcomes by treatment status, where we regard $n,n_0,n_1$ as fixed.

\subsection{The True Variance of the Estimator for the Average Treatment Effect}
\label{subsec:theoretical_variance}
Here we present the true variance of the estimator $\hat\tau_{ATE}$  under random assignment of the treatment. From the $n$ experimental subjects, $n_1$ are selected at random to receive the active treatment, and the remainder are assigned to the control group.
Define
\[ \oy(0)=\frac{1}{N}\sum_{i=1}^N Y_i(0),\hskip1cm \oy(1)=\frac{1}{N}\sum_{i=1}^N Y_i(1),\]
\[ S^2_\ct=\frac{1}{N-1}\sum_{i=1}^N \left(Y_i(0)-\oy(0)\right)^2,\hskip1cm
S^2_\tc=\frac{1}{N-1}\sum_{i=1}^N \left(Y_i(1)-\oy(1)\right)^2,\]
and
\[S^2_{\ct\tc}=\frac{1}{N-1}\sum_{i=1}^N \left(Y_i(1)-Y_i(0)-\tau_{ATE}\right)^2.\]
Then the exact variance of $\hat\tau$, over the randomization distribution, is
\[ \mmv(\hat\tau)=
\frac{S^2_\ct}{n_0}+\frac{S^2_\tc}{n_1}-
\frac{S^2_{\ct\tc}}{N}.\]
See, for example, \cite{ney23}, \cite{AGL14}, \cite{Din17}, and \cite{AAIW17}. %\citet{imbens2015causal}.

\subsection{An Analytical Variance Estimator}
\label{subsec:analytic_variance}
Define
\[ \hat{S}^2_\ct=\frac{1}{n_0-1}\sum_{i=1}^N R_i(1-W_i)\left(Y_i-\oy_\ct\right)^2,\hskip1cm {\rm and}\ \
\hat{S}^2_\tc=\frac{1}{n_1-1}\sum_{i=1}^N R_iW_i\left(Y_i-\oy_\tc\right)^2.\]
Then the standard variance estimator is
\[  \hat\mmv_\neyman:=
\frac{\hat{S}^2_\ct}{n_0}+\frac{\hat{S}^2_\tc}{n_1}.
\]
This estimator ignores the third term in the variance, which is negative, so $\hat\mmv_\neyman$ in general overestimates the true variance. It is possible to give sharp bounds for $S^2_{\ct\tc}$ given the respective marginal distributions of $Y_i(0)$ and $Y_i(1)$. \cite{AGL14} proposed a consistent estimator for the resulting bounds on $\mmv(\hat{\tau})$ that can be expressed as
\[\hat\mmv_\aronow:= \frac{\hat{S}^2_\ct}{n_0}+\frac{\hat{S}^2_\tc}{n_1} - \frac{\underline{\hat{S}}_{\ct\tc}^2}{N}\]
where $\underline{\hat{S}}_{\ct\tc}^2$ is an estimator of the sharp lower bound for $S^2_{\ct\tc}$.\footnote{Such an estimator is
\[\underline{\hat{S}}_{\ct\tc}^2:= \hat{S}^2_\ct + \hat{S}^2_\tc - 2\hat{\sigma}_N^H(y_0,y_1)\]
where $\hat{\sigma}_N^H(y_0,y_1)$ is a consistent estimator for the upper bound for $\cov(Y_i(0),Y_i(1))$, see equation (8) of their paper.}

\subsection{The Classical Bootstrap}

The classical bootstrap corresponds to the case where the uncertainty is purely sampling uncertainty. The bootstrap approximates the cumulative distribution function of the pairs $(Y_i,W_i)$, $F_{YW}(\cdot,\cdot)$ in the population by the empirical distribution function $\hat F_{YW}(\cdot,\cdot)$, where
\[\hat F_{YW}(w,y):=\frac{1}{n}\sum_{i=1}^N R_i\dum\{Y_i\leq y,W_i\leq w\}.\]
It then calculates properties of the estimator given that approximate distribution $\hat F_{YW}(\cdot,\cdot)$. One can interpret the standard bootstrap as imputing all the missing values of $(Y_i,W_i)$ in the population by replicates of the observed values, and thus constructing an artificial population from which we then draw random samples. This perspective is helpful to contrast the different approach underlying the causal bootstrap.

\subsection{The Causal Bootstrap}

\label{subsec:causal_bs_expansion}

Here we initially take the perspective that the uncertainty is solely arising from the stochastic nature of the assignment, as in Table \ref{tabel_assignment}. In the spirit of the above interpretation of the standard bootstrap, we use the observed data to impute all the missing values in the population. Then we simulate the estimator using this partly imputed population.

The difference with the standard bootstrap is in the nature of the missing data process, and how we impute them. Consider unit 1 in Table \ref{tabel_assignment}. In the actual sample this unit receives the active treatment, and so we observe $Y_1(1)$, but we do not know the value of the control outcome for this unit, $Y_1(0)$.

A natural approach is to impute the missing value of $Y_1(0)$ using one of the observed values for $Y_i(0)$, that is, one of the realized values of $Y_i$ for control units. The question is which one to use.
It turns out that it matters how we choose to impute the missing values from the observed values. This issue is related to the term $S_{01}^2$ in the true variance of the estimator $\hat\tau$ for the average treatment effect, the term that is not consistently estimable, and which we typically ignore in practice.

To frame this question, it is useful to start with the joint distribution function of the pairs of potential outcomes in the population,
\[F_{01}^p(y_0,y_1): = \frac{1}{N}\sum_{i=1}^N\dum\left\{Y_i(0)\leq y_0,Y_i(1)\leq y_1\right\}.\]
The average treatment effect, and any other causal parameters of interest, can be written as a functional of this distribution, \[\tau:=\tau(F^p_{01}).\]
Given $F_{01}^p$, the assignment mechanism completely determines the distribution of any estimator, for example the difference in averages by treatment status, $\hat\tau$. This is similar to the way in which in the sampling case knowledge of the joint population distribution allows us to deduce the properties of any estimator.

The problem, and the main difference with the sampling case is that for each unit in the population, at most one of the two potential outcomes $Y_i(0)$ and $Y_i(1)$ is observed so that there is no consistent estimator for $F_{01}^p(\cdot,\cdot)$: In general,  the joint distribution of potential values can be written as
\[F_{01}^p(y_0,y_1) = C(F_0^p(y_0),F_1^p(y_1))\]
where the copula $C:[0,1]^2\mapsto[0,1]$ is a function that is nondecreasing in either argument for each value of $x$. By Sklar's theorem (e.g. stated as Theorem 2.3.3 in \cite{Nel06}), such a copula exists even though it need not be unique unless the marginal distributions $F_0^p,F_1^p$ are continuous. In the following, we let
\[\mathcal{C}:=\left\{C:[0,1]^2\mapsto[0,1],\; C(u,v) \textnormal{ nondecreasing in $u$ and $v$}\right\}\]
denote the set of all possible copulae.

It is important to note that although the marginal distributions $F_0^p,F_1^p$ can be estimated consistently from a completely randomized experiment as sample size grows, the data on realized treatments and outcomes impose no empirical restrictions on the copula $C(u,v)$ for the joint distribution of $(Y_i(0),Y_i(1))$. Hence, neither the parameter $\tau(F_{01}^s)=\tau(C(F_0^s,F_1^s))$ nor the distribution of an estimator $\hat{\tau}$ need in general not be point-identified.

In the spirit of the variance estimator in \cite{AGL14}, we address this challenge by simulating the distribution of $\hat{\tau}$ using an estimator for the population distribution $F_{01}^p$ that is \emph{conservative} with respect to the copula in a sense to be made more precise below. To illustrate the broader conceptual idea, consider an estimator \[\hat{\tau}:=\tau(\hat{F}_{0},\hat{F}_{1})\] for a general functional $\tau(F_{01})$ of the distribution of potential values. Under regularity conditions,\footnote{See e.g. \cite{BGo01} for regularity conditions for finite-population expansions of this type.} such an estimator admits a stochastic expansion of the form
\[\hat{\tau}-\tau(F_{01}^p) = \mu(F_{01}^p) + n^{-1/2}\sigma(F_{01}^p)Z + n^{-1}\kappa(F_{01}^p) + o_P(n^{-1})\]
where $Z\sim N(0,1)$. The first-order ``bias" term \[\mu(F_{01}^p):=\mathbb{E}_{F_{01}^p}[\hat{\tau}] - \tau(F_{01}^p)\] and the scale parameter \[\sigma^2(F_{01}^p):=\lim_{N}n\var_{F_{01}}(\hat{\tau})\] are deterministic functions of the unknown distribution $F_{01}^p=C(F_0^p,F_1^p)$, and the limit for the asymptotic variance is taken as $N$ and $n:=n_N$ grow large. The second-order approximation error $\kappa(F_{01}^p)$ is a tight random variable whose distribution also depends on $F_{01}^p$.

If the functional $\tau(F_{01}^p)$ is not point-identified, then $\mu(F_{01}^p)$ may take values in a set whose bounds may be characterized in terms of the marginal distributions $F_0^p,F_1^p$. Specifically, given the marginal distributions $F_0^p,F_1^p$ we have sharp bounds of the form
\[\mu_L(F_0^p,F_1^p):=\inf_{\tilde{C}\in\mathcal{C}}\mu(\tilde{C}(F_0^s,F_1^s)\leq \mu(F_{01}^s)\leq \sup_{\tilde{C}\in\mathcal{C}}\mu(\tilde{C}(F_0^s,F_1^s))=:\mu_U(F_0^p,F_1^p),\]
that are generally available, see e.g. \cite{HSC97} and \cite{Man97}. Similarly we can form bounds for the variance,
\[\sigma_L(F_0^p,F_1^p):=\inf_{\tilde{C}\in\mathcal{C}}\sigma(\tilde{C}(F_0^s,F_1^s)\leq \sigma(F_{01}^s)\leq \sup_{\tilde{C}\in\mathcal{C}}\sigma(\tilde{C}(F_0^s,F_1^s))=:\sigma_U(F_0^p,F_1^p)\]

For a given inference problem, the bootstrap has to estimate these quantities conservatively with respect to the unknown copula $C(\cdot)$, which can be done iteratively as follows: we first need to determine which couplings $C_0^*$ attain the value of $\mu(C_0^*(F_0,F_1))$ which is least favorable for the inference problem at hand. Within the (not necessarily singleton) set $\mathcal{C}_0^*$ of such couplings, we then determine the least-favorable value of $\sigma(C_1^*(F_0,F_1))$ for $C_1^*\in\mathcal{C}_0^*$. We can apply this principle recursively either until the resulting set $\mathcal{C}_k^*$ contains a unique copula, or until we reach the order of approximation desired for formal results regarding the bootstrap procedure. This results in an estimate $\hat{F}_{01}^*:=C_k^*(\hat{F}_{0},\hat{F}_{1})$ for the population distribution $F_{01}^p$ that is conservative regarding the inference task at hand. The causal bootstrap then approximates the distribution of $\hat{\tau}$ by sampling and randomization from a population $\hat{F}_{01}^*$ using the known sampling and assignment mechanism.

\subsection{Least Favorable Coupling for the Average Treatment Effect}

\label{sec:SATE_section}

In this paper, we consider the special case of two-sided confidence intervals based on a t-ratio for the sample average treatment effect. The case of the average treatment effect has been the main focus of the previous literature. It is a special case for our problem in that the copula does not matter for estimation - by inspection, the functional
\[\tau_{ATE}(F_{01})=\mathbb{E}_{F_{01}}[Y_i(1)]-\mathbb{E}_{F_{01}}[Y_i(0)]=\mathbb{E}_{F_{1}}[Y_i(1)]-\mathbb{E}_{F_{0}}[Y_i(0)]=:\tau(F_0,F_1)\]
does not depend on the copula, and the default estimator
\[\hat{\tau}_{ATE}:=\tau(\hat{F}_{0},\hat{F}_{1})\equiv\frac1{n_1}\sum_{i=1}^NR_iW_iY_i - \frac1{n_0}\sum_{i=1}^NR_i(1-W_i)Y_i\]
is known to be unbiased for $\tau(C(F_0,F_1))$ under any coupling so that $\mu(C(F_0,F_1))\equiv0$ for each $C$. In order to ensure that the estimand is well-defined and satisfies other regularity conditions for the bootstrap, we make the following assumptions:

\begin{ass}\label{SATE_mom_ass} The first four moments of $F_0(y_0)$ and $F_1(y_1)$ are bounded.
\end{ass}

Also for a two-sided confidence interval constructed from inverting a t-test based on $\hat{\tau}_{ATE}$, the least favorable coupling must attain the upper bound for the asymptotic variance,
\[\sigma_U^2(F_0,F_1)=\sup_{C\in\mathcal{C}}\sigma^2(C(F_0,F_1))=:\sigma^2(F_0,F_1)\]
We next show that $\sigma^2(C(F_{0},F_1))$ is uniquely maximized at the joint distribution corresponding to the isotone assignment which matches values of $Y_i(0)$ to values of $Y_i(1)$ while preserving their respective marginal distributions. More formally, the joint distribution of the potential outcomes under the isotone coupling is characterized by the copula
\[C^{iso}(u,v):=\min\{u,v\}\]
We find that the upper bound on the variance is in fact uniquely attained at the isotone coupling. Therefore an estimator for the distribution of $\hat{\tau}_{ATE}$ which assumes the isotone coupling is asymptotically conservative at any order of approximation.

\begin{prp}\textbf{(Least Favorable Coupling for the ATE).}\label{sate_iso_prp} Suppose that Assumption \ref{SATE_mom_ass} holds. Then, given the marginal distributions $F_0,F_1$, the variance bound is uniquely attained at
\[\sigma^2(F_0,F_1):=\lim_Nn\var_{F_{01}^{iso}}(\hat{\tau})\]
where $F_{01}^{iso}:=C^{iso}(F_0,F_1)$ is the joint distribution corresponding to the isotone coupling.
\end{prp}

The fact that the variance bound is attained at the isotone coupling is widely known (see e.g. \cite{Bec73}, \cite{FPa10}, \cite{Sto10}, and \cite{AGL14}), for expositional purposes we provide a proof in the appendix. We establish the slightly stronger conclusion that the distribution under the isotone coupling is in fact maximal with respect to second-order stochastic dominance. For our approach it is also important to establish that this maximum is unique in the sense that the joint distribution resulting from any other coupling yields a variance that is strictly lower than $\sigma_U^2(F_0,F_1)$. In particular, for confidence intervals based on the Gaussian asymptotic distribution, the isotone coupling does indeed constitute the least favorable coupling.

We also want to stress that there are other causal estimands of interest (including the distribution of treatment effects and its quantile) for which the isotone assignment is not the least favorable coupling (see e.g. \cite{HSC97}, \cite{FWu10}, \cite{FPa10}, \cite{LDD18}). \cite{CSS76} give conditions under which the isotone assignment does in fact constitute the least favorable bound for a functional of the joint distribution.

\subsection{Related Literature}

Worst case bounds on the distributions of potential outcomes and treatment effects and their quantiles have been analyzed by \cite{HSC97}, \cite{Man97}, \cite{FRi08}, \cite{FPa10}, \cite{FPa12}, \cite{FWu10}, and \cite{LDD18}. This literature uses theoretical results on dependency bounds for functions of several random variables which were developed among others by \cite{CSS76}, \cite{Mak82}, \cite{FNS87}, and \cite{WDo90}. \cite{Sto10} establishes that a class of spread parameters is monotone with respect to conventional stochastic orders of distribution, and shows how to derive parameter bounds for causal inference. Several of these studies also propose inference procedures that account for sampling uncertainty rather than randomization error. In contrast, for our problem we need to explicitly construct the respective couplings that achieve the lower and upper bounds to the parameter, and in addition the largest randomization variance for an estimator of either bound.

\cite{Rob88} proposes a confidence interval for a causal parameter based on the least-favorable coupling for a binary outcome variable. \cite{AGL14} propose an estimator of the sharp upper bound for the randomization variance of the average treatment effect in completely randomized experiments. Our approach of embedding the finite-population randomization distribution into an asymptotic sequence of sampling experiments closely follows \cite{AAIW17}. Our results make use of a finite-population CLT for the empirical process developed by \cite{Bic69} for the two-sample problem. Finite-sample central limit theorems for randomization inference were also provided by \cite{LDi17}. Bootstrap methods for sampling from finite populations (without replacement) have been proposed by \cite{BFr84} and \cite{BBH94}. For this problem the main challenge in generating the finite bootstrap population is that the size of the super-population $N$ may be a non-integer multiple of $n$. We propose a new alternative for estimating the potential outcome distribution for a super-population of exact size $N$ and for which the marginal distributions coincide with their empirical analogs up to rounding error.

\subsection{Comparison to Fisher's Exact Test}
\label{subsec:exact_test_comp}

Bootstrap inference on the average treatment effect as proposed in this paper bears some conceptual similarities with Fisher's exact test of the sharp null of no unit-level treatment effect (see e.g. \cite{Ros02}, \cite{IRu15}, \cite{Din17}), $Y_i(0)=Y_i(1)$ with probability 1. One important distinction is that the justification for our procedure is only asymptotic, whereas the Fisher exact test is valid in finite samples.

Furthermore, the Fisher exact test evaluates the randomization distribution of the estimated ATE under the sharp null of no or a constant unit-level treatment effect. The sharp null not only implies that the joint distribution of $Y_i(0)$ and $Y_i(1)$ corresponds to the isotone assignment, but also equality of the marginal distributions $F_0(y)=F_1(y)$, which may in fact be rejected by the data under the null of a zero \emph{average} treatment effect. In that case even a conservative estimator of the randomization variance may in fact be smaller than that implied by zero, or constant, unit-level effects. More generally, when Fisher's sharp null fails and $F_0(y)\neq F_1(y)$, the bootstrap estimate of the randomization variance can in several important scenarios be smaller than that implicit in Fisher's exact test, in which case our procedure is asymptotically more powerful.

Specifically, standard variance calculations (see e.g. \cite{Din17}) imply that the implicit variance estimate for Fisher's exact test under the null of no average effect is \[\mmv_{Fisher}(\hat{\tau}_{ATE})=\var(Y_i)\left(\frac1{n_1}+\frac1{n_0}\right)=\frac{n_0S^2_\ct + n_1S^2_\tc}{n}\left(\frac1{n_1}+\frac1{n_0}\right).\] We can compare this to the actual variance stated in Section \ref{subsec:theoretical_variance},
\[ \mmv(\hat\tau)=
\frac{S^2_\ct}{n_0}+\frac{S^2_\tc}{n_1}-
\frac{S^2_{\ct\tc}}{N}.\]

Our bootstrap procedure implies a conservative estimate, i.e. a sharp lower bound for $S^2_{\ct\tc}$ from the isotone coupling of the potential outcomes, which is strictly positive whenever the marginal distributions of $Y_i(0)$ and $Y_i(1)$ are not the same. The comparison between the terms $\frac{S^2_\tc}{n_1} + \frac{S^2_\ct}{n_0}$ and $\var(Y_i)\left(\frac1{n_1}+\frac1{n_0}\right)$ is generally ambiguous - \cite{Din17} describes several cases in which the randomization variance implied by Fisher's test is strictly larger, and his conclusions carry over to the bootstrap procedure in this paper. On the other hand it is important to note that when $\frac{S^2_\tc}{n_1} + \frac{S^2_\ct}{n_0}>\var(Y_i)\left(\frac1{n_1}+\frac1{n_0}\right)$, Fisher's exact test over-rejects under the null hypothesis of no average treatment effect, so that this potential power advantage for the Fisher test only arises in situations in which the exact test does not provide a valid test of that null. We illustrate this possibility using Monte Carlo simulations in Section \ref{sec:monte_carlo}.

The relationship between Fisher's sharp null and Neyman's null hypothesis of no average effect is clarified in \cite{Din17}, who also shows that Neyman's test of the null of no average effect is weakly more powerful against alternatives than Fisher's exact test. Fisher's exact null also implies that the distribution of $\Delta_i$ is degenerate at a constant, however the power comparison for the ATE does not carry over to set-identified objects like quantiles or the c.d.f. of $\Delta_i$ since the bounds for the identified set are typically not attained at the isotone coupling that is implied by the sharp null. %[does this make sense if the randomization distribution already imposes a constant effect? - I guess it depends on what functional of the distribution is evaluated]

\section{Bootstrap Procedure}

This section describes the bootstrap procedure for confidence intervals for the average treatment effect, in which case the least favorable coupling is the isotone (rank-preserving) assignment by Proposition \ref{sate_iso_prp}. The method allows for sampling and randomization uncertainty, where we consider a sampling experiment under which the researcher observes $n$ units that are selected at random out of a population of $N$ units. For the purposes of asymptotic approximations, we assume that the population of interest in turn consists of $N$ i.i.d. draws from an encompassing distribution $F_{01}$.

\begin{ass}\label{density_ass}\textbf{(Sampling Experiment)} The population consists of $N$ units with potential values $\left(Y_i(0),Y_i(1)\right)_{i=1}^N$ which are i.i.d. draws from the distribution $F_{01}(y_0,y_1)$. The $n$ observed units are sampled at random and without replacement from the population,
\[Y_i(0),Y_i(1)\ind R_i\]
where we denote $q:=\frac{n}N\in(0,1]$.
\end{ass}

We assume throughout that the treatment $W_i\in\{0,1\}$ is binary, and that the outcome $Y_i(W_i)$ for unit $i$ does not vary with the treatment status assigned to other units. The latter requirement is also known as individualistic treatment response, or Stable Unit Treatment Value Assumption (SUTVA). We assume furthermore that the experiment is completely randomized:

\begin{ass}\label{unconfoundedness_ass}\textbf{(Complete Randomization)} Treatment assignment is completely randomized, that is for each unit with $R_i=1$ we have
\[(Y_i(0),Y_i(1))\ind W_i\]
where $W_i=1$ for $n_1$ units selected at random and without replacement from the $n$ observations with $R_i=1$, and the propensity score $p:=\frac{n_1}n$ satisfies $0<p<1$.
\end{ass}

For greater clarity of exposition we also assume that the researcher observes no further covariate information. The approach of this paper can be generalized to observational studies under unconfoundedness, and experiments with imperfect compliance for which unconfoundedness fails, but intention to treat is (conditionally) independent of potential outcomes and can serve as an instrumental variable to identify causal effects on a population of compliers.

Given a sample generated according to Assumptions \ref{density_ass} and \ref{unconfoundedness_ass}, we denote the point estimate for the average treatment effect
\[\hat{\tau}:=\tau(\hat{F}_{0},\hat{F}_{1})\]
and the upper variance bound
\[\hat{\sigma}:=\sigma(\hat{F}_{0},\hat{F}_{1})\]
For the purposes of this paper, the main target of interest for the causal bootstrap is the distribution of the t-ratio
\[T:=\sqrt{n}\frac{\hat{\tau}-\tau}{\hat{\sigma}}\]

%In a companion paper we will discuss adaptations to our approach to observational studies under unconfoundedness, and experiments with imperfect compliance for which unconfoundedness fails, but intention to treat is (conditionally) independent of potential outcomes and can serve as an instrumental variable to identify causal effects on a population of compliers. We also discuss challenges in extending our approach when treatment assignment is not independent across units.

\subsection{Bootstrap Algorithm}

\label{sec:algorithm_sec}

The proposed bootstrap algorithm proceeds in four main steps:
\begin{enumerate}
\item We obtain nonparametric estimates of the potential outcome distributions $F_0(y_0)$ and $F_1(y_1)$ from the units for which $W_i=0$ ($W_i=1$, respectively) in the actual experiment.
\item We create an empirical population of size $N$, $\left(\tilde{Y}_i,\tilde{W}_i\right)_{i=1}^N$ by generating an appropriate number of replicas of the sample of $n$ draws for $W_i,Y_i$. If the sample is the population, $n=N$, we can skip this step.
\item We then impute potential values $\tilde{Y}_i(0),\tilde{Y}_i(1)$ for each unit $i=1,\dots,N$, where $\tilde{Y}_i(\tilde{W}_i) = \tilde{Y}_i$ and $\tilde{Y}_i(1-\tilde{W}_i)$ is obtained from the estimated potential outcome distributions and the least-favorable copula for the parameter of interest.
\item Finally, we simulate the randomization distribution by repeatedly drawing $n$ units $Y_i^*(0),Y_i^*(1)$ out of that empirical population without replacement and generating randomization draws $W_1^*,\dots,W_n^*$. We then evaluate the sample average treatment effect for the bootstrap sample $\left(Y_i^*(W_i^*),W_i^*\right)_{i=1}^n$ obtained using the imputed potential outcomes.
\end{enumerate}
%Note that the respective least-favorable copula, and therefore the imputed values for evaluating the lower bound is different from that for the upper bound, so that this procedure has to be repeated separately for either of the two bounds.

Given the simulated randomization distribution for the estimated bounds, we can estimate the percentiles of the t-ratios that are needed to construct confidence intervals for the functional. We next describe each of these steps in greater detail.

\subsection{Generating the Empirical Population}

To obtain the empirical population of size $N$, we generate replicates of the $n$ observed units, however not necessarily of the same number for each observation when $N$ is not an integer multiple of $n$. We propose the following procedure for doing so:

\begin{itemize}
%\item We generate an empirical rank $\hat{\tau}_i:=\hat{F}_{W_in}(\tilde{Y}_i)$.
%\[\hat{\tau}_i:=\left\{\begin{array}{lcl}\hat{F}_{0}(Y_i)&\hspace{0.3cm}&\textnormal{if }W_i=0\\
%    \hat{F}_{1}(Y_i)&&\textnormal{if }W_i=1\end{array}\right.\]
%    for each observed unit $i=1,\dots,n$.
\item We create the samples $\left(Y_j^0,\hat{U}_j^0\right)_{j=1}^{n_0}$ of values for $Y_i$ for the $n_0$ units with $W_i=0$, and $\left(Y_j^1,\hat{U}_j^1\right)_{j=1}^{n_1}$ with values $Y_i$ for the $n_1$ units with $W_i=1$. We assume that each sample is ordered, $Y_k^w\leq Y_{k+1}^w$ for all $k$, and the rank variable $\hat{U}_j^w=\frac{j}{n_w}$ for $w=c,t$.
%$\left(Y_{j:n_0}^0,\hat{\tau}_{(j)}\right)_{j=1}^{n_0}$ for the $n_0$ units with $W_i=0$ and $\left(Y_{j:n_1}^1,\tilde{W}_{(j)},\hat{\tau}_{(j)}\right)_{j=1}^{n_1}$ for the $n_1$ units with $W_i=1$, with indices ordered by $\hat{\tau}_{(j)}$, i.e. $\hat{\tau}_{(1)}\leq\hat{\tau}_{(2)}\leq\dots\leq\hat{\tau}_{(\max\{n_0,n_1\})}$.
\item Let $N_0=\lceil\frac{n_0}nN\rceil$ and $N_1=N-N_0$. We generate the empirical population $\left(\tilde{Y}_i,\tilde{W}_i\right)_{i=1}^N$ by including $M_{j}^0:=\lceil \hat{U}_{j+1}^0N_0\rceil - \lceil\hat{U}_j^0N_0\rceil$
    copies of $Y_j^0$ with $\tilde{W}_j=0$ and $M_{j}^1:=\lceil\hat{U}_{j+1}^1N_1\rceil - \lceil\hat{U}_j^1N_1\rceil$ copies of $Y_j^1$ with $\tilde{W}_j=1$.
\end{itemize}

Since the respective maxima of $\hat{U}_j^0,\hat{U}_j^1$ are equal to $1$ for either of the two strata (corresponding to $W_i=0$ and $W_i=1$, respectively), $\sum_{j=1}^n((1-\tilde{W}_{(j)})M_{(j)}^0+\tilde{W}_{(j)}M_{(j)}^1)=\lceil N_0\rceil + \lceil N_1\rceil=N$ so that this procedure ensures that the empirical population has size equal to $N$. Also, for $n$ and $N$ sufficiently large, the respective empirical distributions of $\tilde{Y}_i$ among units with $\tilde{W}_i=0$ and $\tilde{Y}_i$ among units with $\tilde{W}_i=1$ are, up to an approximation error of the order $n^{-1}$, equal to $\hat{F}_{0}$ and $\hat{F}_{1}$, respectively.

\subsection{Imputing Missing Counterfactuals}

For the specific case of two-sided inference for the average treatment effect, Proposition \ref{sate_iso_prp} shows that the least favorable coupling corresponds to the isotone assignment $C^{iso}(u,v):=\min\{u,v\}$. For other inference problems, the missing counterfactuals would have to be imputed by drawing from the appropriate least-favorable coupling, following the strategy outlined in Section \ref{subsec:causal_bs_expansion}.

In order to generate an empirical population with joint distribution $\hat{F}_{01}^{iso}:= C^{iso}(\hat{F}_{0},\hat{F}_{1})$, we can simply impute the missing counterfactuals according to:
\begin{eqnarray}
\nonumber \tilde{Y}_i(0)&:=&\left\{\begin{array}{lcl}\tilde{Y}_i&\hspace{0.3cm}&\textnormal{if }\tilde{W}_i=0\\
\hat{F}_{0}^{-1}\left(\hat{F}_{1}(\tilde{Y}_i)\right)&&\textnormal{otherwise}\end{array}\right.\\
\label{impute_pot_outcome}\tilde{Y}_i(1)&:=&\left\{\begin{array}{lcl}\tilde{Y}_i&\hspace{0.3cm}&\textnormal{if }\tilde{W}_i=1\\
\hat{F}_{1}^{-1}\left(\hat{F}_{0}(\tilde{Y}_i)\right)&&\textnormal{otherwise}\end{array}\right.
\end{eqnarray}
For functionals $\tau(F_{01})$ of the potential outcome distribution other that the average treatment effect, or inference problems other than two-sided confidence intervals, the least favorable coupling will be of a different form, so this step would have to be replaced by a procedure imputing the missing counterfactuals from a different coupling.

%In the specific cases discussed in this paper the least-favorable copula is always degenerate in the sense that it represents a deterministic assignment between quantiles of the distributions of $Y_i(0)$ and $Y_i(1)$, respectively. In the general case a potential concern could be that the imputation scheme involves the random rank variables $U_i$ and is therefore non-deterministic. However, this type of noise in sampling from the distribution $F_{01}(y_0,y_1)$ vanishes in the limit and therefore does not pose any problems for the asymptotic properties of the procedure. Also, the inversions of $C^*(u,v)$, $\hat{F}_{0}(y_0)$ and $\hat{F}_{1}(y_1)$ in computing the imputed counterfactuals may not be straightforward in the general case, but for average, c.d.f. and quantiles of the treatment effect distribution we have closed-form representations and we give algorithms for generating these values.

\subsection{Resampling Algorithm}

% recompute functional give sample

For the $b$th bootstrap replication, we initially draw $n$ units $\left(Y_{ib}^*(0),Y_{ib}^*(1)\right)$ from the empirical population at random and without replacement.

Given a known propensity score $p:=P(W_i|R_i=1)$, for the $b$th bootstrap replication we can generate $W_{1b}^*,\dots,W_{nb}^*$ as independent Bernoulli draws with success probability $P(W_{ib}^*=1)=p$ and obtain the bootstrap sample $Y_{1b}^*,\dots,Y_{nb}^*$, where $Y_{ib}^*:=Y_{ib}^*(W_{ib}^*)$.

We can then compute the bootstrap analogs of the estimated c.d.f.s $\hat{F}_{0b}^{*}(y_0):=\frac1{n_0}\sum_{i=1}^NR_{ib}^*(1-W_{ib}^*)\dum\{Y_{ib}^*\leq y_0\}$ and $\hat{F}_{1b}^{*}:=\frac1{n_1}\sum_{i=1}^NR_{ib}^*W_{ib}^*\dum\{Y_{ib}^*\leq y_1\}$, the corresponding estimates of the average treatment effect, and the variance bound,
\begin{eqnarray}
\nonumber \hat{\tau}_{b}^*:=\tau(\hat{F}_{0b}^{*},\hat{F}_{1b}^{*})\\
\nonumber \hat{\sigma}_{b}^*:=\sigma(\hat{F}_{0b}^{*},\hat{F}_{1b}^{*})
\end{eqnarray}

We then record the studentized values of the bootstrap estimates,
\[T_{b}^*:=\sqrt{n}\frac{\hat{\tau}_{b}^*-\hat{\tau}}{\hat{\sigma}_b^*}\]
Repeating the resampling step $B$ times, we obtain a sample $(T_{1}^*,\dots,T_{B}^*)$ that constitutes independent draws from the bootstrap estimator of the randomization distribution and can be used to construct critical values for tests or confidence intervals.

\subsection{Confidence Intervals}

We consider confidence intervals constructed by inverting a t-test based on the point estimate $\hat{\tau}:=\tau(\hat{F}_{0},\hat{F}_{1})$ and given the variance bound $\hat{\sigma}:=\sigma(\hat{F}_{0},\hat{F}_{1})$ introduced before. The proposed confidence intervals for $\tau$ are then of the form
\begin{equation}\label{confidence_interval}CI(1-\alpha):=\left[\hat{\tau} - n^{-1/2}\hat{\sigma}\hat{c}(1-\alpha),
\hat{\tau} - n^{-1/2}\hat{\sigma}\hat{c}(\alpha)\right] \end{equation}
We use bootstrap approximations to the randomization distribution of the t-ratio $n^{1/2}(\hat{\tau}-\tau)/\hat{\sigma}$ under the least favorable coupling in order to determine the critical values. Specifically, let $\hat{G}(z):=\frac1B\sum_{b=1}^B\dum\{T_{b}^*\leq z\}$ denote the empirical distribution for the bootstrap samples obtained from the previous step. We then estimate the critical values using
$\hat{c}(\alpha):=\hat{G}^{-1}(\alpha)$ and $\hat{c}(1-\alpha):=\hat{G}^{-1}(1-\alpha)$.

\section{Monte Carlo Simulations}

\label{sec:monte_carlo}

%We next briefly illustrate this in a simple setting, and compare it to the standard bootstrap.

We next compare the performance of this causal bootstrap with the standard bootstrap and other alternative methods based on sampling or randomization designs. Specifically, we consider confidence intervals using Gaussian critical values with the respective analytic estimators of the sampling variance $\hmmv_\neyman$ and the causal variance $\hmmv_\aronow$ given in Section \ref{subsec:analytic_variance}. We also consider Gaussian inference using the variance estimators $\hmmv_\sboot$ and $\hmmv_\cboot$ obtained from the classical (sampling) bootstrap, and the causal bootstrap proposed in this paper. We compare these to confidence intervals from inverting Fisher's exact test, and confidence intervals from the standard and the causal bootstrap for the t-statistic based on either sampling or causal variance estimate. Throughout we will restrict our attention to the case $n=N$, i.e. when the full population of interest is observed.

%\begin{enumerate}
%\item the Gaussian distribution, with the variance $\hmmv_\neyman$,
%\item the Gaussian distribution, with the variance $\hmmv_\aronow$,
%\item the Gaussian distribution, with the variance $\hmmv_\sboot$,
%\item the Gaussian distribution, with the variance $\hmmv_\cboot$,
%\item the Fisher (exact) randomization test,
%\item the standard bootstrap for the t-statistic, with the variance $\hmmv_\neyman$,
%\item the standard bootstrap for the t-statistic, with the variance $\hmmv_\aronow$,
%\item the causal bootstrap for the t-statistic, with the variance $\hmmv_\neyman$, and
%\item the causal bootstrap for the t-statistic, with the variance $\hmmv_\aronow$.
%\end{enumerate}

%From the theoretical results in this paper, we would expect that
%\begin{enumerate}
%  \item With $n_0$, $n_1$ large and a constant treatment effect, all methods work well.
%  \item With $n_0$, $n_1$ small and a constant treatment effect, normal distribution based methods do not work well compared to pivotal %statistic based bootstrap methods.
%  \item With $n_0$, $n_1$ large, big difference between Aronow and Neyman variance, standard methods including bootstrap do not work well.
%\item With $n_0$, $n_1$ small,  big difference between Aronow and Neyman variance, only pivotal causal bootstrap should work well.
%\end{enumerate}

We first consider three different simulation designs to illustrate the main points of comparison between the causal bootstrap and the main alternatives for causal inference.
\begin{itemize}
\item Design I sets $n_0=n_1=100$ and draws potential outcomes according to $Y_i(0)\sim\mathcal{N}(0,1)$, $Y_i(1)=Y_i(0)$. In this setting, treatment effects are constant at $Y_i(1)-Y_i(0)\equiv0$ and the marginal distributions $F_0^p(y)\equiv F_1^p(y)$, so that all procedures should be expected to do well.
\item For Design II we again have $n_0=n_1=100$, but generate potential outcomes as $Y_i(0)\sim\mathcal{N}(0,1)$, and $Y_i(1)=0$. In that case, the marginal distributions $F_0^p(y)$ and $F_1^p(y)$  are different, so that causal standard errors and the causal bootstrap should do better than their sampling analogs.
\item Design III replicates Design II at a smaller sample size, where $n_0=n_1=20$, and $Y_i(0)\sim\mathcal{N}(0,1)$, $Y_i(1)=0$.
\item For Design IV, $n_0=n_1=20$, and we generate non-Gaussian potential outcomes where $Y_i(1)=0$ and $Y_i(0)$ is a mixture that is drawn from $\mathcal{N}(0,1)$ with probability 0.9, and from $\mathcal{N}(0,16)$ with probability 0.1. This design highlights the difference between the bootstrap and Gaussian inference, which is no longer exact for this design.
\end{itemize}
Simulation results are shown in Table \ref{tab:des123_sim}, where we compare coverage rates of nominal 95\% confidence intervals, and the corresponding standard errors for each of the three designs. If a particular method does not directly calculate standard errors, we calculate the standard errors by taking the ratio of the difference between the upper and lower limit of the confidence interval and dividing by 2 times 1.96. We also report the nominal confidence level and theoretical causal standard error $\sqrt{\mmv(\hat\tau)}$ for each design in the last row (``Target"). With the exception of Fisher's exact test, all methods rely on asymptotics, and should therefore not be expected to achieve exact coverage at the nominal level.

Under the first design, causal and sampling standard errors coincide, and the exact distribution of $\hat{\tau}_{ATE}$ is Gaussian. Furthermore, Fisher's exact null holds true, so all methods should work well. Under the second design, the exact distribution of $\hat{\tau}_{ATE}$ is again Gaussian but a conservative estimate for the variance $\var(Y_i(1)-Y_i(0))$ is strictly positive, so the causal standard error is strictly smaller than the sampling-based standard error. Hence for Design II, inference based on sampling based standard errors or the standard bootstrap should be expected to be conservative, whereas inference using causal standard errors or the causal bootstrap may still be conservative but coverage should be closer to the nominal level than for sampling-based methods.

Design III repeats Design II for a smaller sample size, which reveals a modest downward bias in the (bootstrap or analytical) causal standard errors, resulting in rejection rates exceeding the nominal level. Such a bias should be expected since the causal standard error corresponds to the plug-in estimator $\sigma(\hat{F}_0,\hat{F}_1)$, where under regularity conditions $\sigma(F_0,F_1)$ is a smooth but nonlinear functional of the marginal distributions $F_0,F_1$. While the bootstrap should not be expected to provide refinements for estimating the standard error (a non-pivotal quantity), a refinement for inference based on the studentized estimator corrects for the leading term of that bias and therefore result in rejection rates closer to the desired level.

Design IV has heterogeneous treatment effects and non-Gaussian marginal distributions for potential outcomes, where the tails of the marginal distribution of $Y_i(0)$ are thicker than for the Gassian distribution. In this setting sampling-based methods should be more conservative than their causal analogs, and the pivotal bootstrap may provide refinements over Gaussian inference with causal standard errors. Since sample size for the third design is fairly small and all methods except for Fisher's exact test rely on asymptotics, all inference methods exhibit modest size distortions. However simulated coverage rates for the pivotal bootstrap are close to the nominal level, and simulation results in Tables \ref{tab:coupling_sim} and \ref{tab:refinements_sim} confirm that coverage rates approach the desired nominal level when sample sizes get sufficiently large.

\begin{table}[ht]\label{tab:des123_sim}
\caption{95\% Confidence Intervals And Standard Errors}
\begin{center}
\scriptsize{\begin{tabular}{cccccccccccccccc}
&&&&\multicolumn{2}{c}{Design I}&&\multicolumn{2}{c}{Design II}&&\multicolumn{2}{c}{Design III}&&\multicolumn{2}{c}{Design IV}\\[2pt]
 Variance &\multicolumn{1}{c}{Bootstrap}&Pivotal&&Cov & Med&&Cov & Med&&Cov & Med&&Cov & Med\\
  Estimator &Version& Statistic &\hspace{0.3cm}&  Rate &  s.e.&&  Rate &  s.e.&&  Rate &  s.e.&&  Rate &  s.e.\\[4pt]\cline{5-6}\cline{8-9}
  \cline{11-12}\cline{14-15}\\
 $\hmmv_\neyman$ &N/A& No                        &&    0.9536 &   0.1412 &&   0.9950  &  0.0999 &&   0.9870 &   0.2218  &&  0.9776  &  0.3330 \\
  $\hmmv_\aronow$ &N/A&No                        &&    0.9528 &   0.1404 &&   0.9524  &  0.0706 &&   0.9334 &   0.1568  &&  0.9116  &  0.2354 \\
 $\hmmv_\sboot$ &Standard&No                     &&    0.9518 &   0.1405 &&   0.9944  &  0.0994 &&   0.9850 &   0.2162  &&  0.9744  &  0.3245 \\
 $\hmmv_\cboot$ &Causal&No                       &&    0.9512 &   0.1400 &&   0.9494  &  0.0704 &&   0.9302 &   0.1548  &&  0.9084  &  0.2325 \\[4pt]
 \multicolumn{3}{c}{Fisher's Exact Test}         &&    0.9766 &   0.1411 &&   0.9630  &  0.0999 &&   0.9626 &   0.2219  &&  0.9698  &  0.3332 \\[4pt]
 $\hmmv_\neyman$ &Standard&Yes                   &&    0.9534 &   0.1421 &&   0.9954  &  0.1012 &&   0.9900 &   0.2404  &&  0.9838  &  0.3865 \\
  $\hmmv_\aronow$ &Standard&Yes                  &&    0.9528 &   0.1433 &&   0.9954  &  0.1012 &&   0.9900 &   0.2404  &&  0.9838  &  0.3865 \\
 $\hmmv_\neyman$ &Causal& Yes                    &&    0.9526 &   0.1414 &&   0.9510  &  0.0715 &&   0.9446 &   0.1681  &&  0.9434  & 0.2802 \\
  $\hmmv_\aronow$ &Causal&Yes                    &&    0.9530 &   0.1419 &&   0.9510  &  0.0715 &&   0.9446 &   0.1681  &&  0.9434  &  0.2802 \\[4pt]
&\vspace{-0.2cm}\\\cline{5-6}\cline{8-9} \cline{11-12}\cline{14-15}\\
\multicolumn{3}{c}{Target}                       &&    0.9500 &   0.1414 &&   0.9500  &  0.0707 &&   0.9500 &   0.1581  &&  0.9500  &  0.2500 \\[1pt]

&\vspace{-0.2cm}\\\cline{5-6}\cline{8-9} \cline{11-12}\cline{14-15}\\
\end{tabular}}		\end{center} \end{table}

% (a) balanced non-Gaussian design to show refinements - 50 100 200 400
% (b) unbalanced Gaussian design to show challenges for Fisher's test 20/50; 80/200

We next illustrate the role of the coupling of the potential values where we draw $(Y_i(0),Y_i(1))$ from a bivariate Gaussian distribution with variances $\var(Y_i(0))=0.5$ and $\var(Y_i(1))=2$ and correlation coefficient of the two potential values, $\varrho_{01}\in\{-1,0,1\}$. From our theoretical results, we should expect Gaussian inference using causal standard errors and the causal bootstrap to have asymptotically exact coverage under the isotonic coupling $\varrho_{01}=1$ and be conservative when $\varrho_{01}<1$. Furthermore, for any coupling this design implies heterogeneous treatment effects, so that Fisher's exact test does not in general control nominal confidence size for the average treatment effect. Given the calculations in Section \ref{subsec:exact_test_comp} we designed the experiment deliberately to illustrate the potential of Fisher's exact procedure to underestimate the spread of the randomization distribution, where $n_0>n_1$ and $\var(Y_i(1))>\var(Y_i(0))$. Since the potential outcomes follow a Gaussian distribution, we should not expect refinements for the bootstrap relative to Gaussian inference.

\begin{table}\label{tab:coupling_sim}\caption{Coverage of nominal 95\% Confidence Intervals, Gaussian Potential Outcomes with Different Couplings}
\begin{center}
\hspace*{-0.9cm}
\scriptsize{\begin{tabular}{ccccccccccccccc}

  Variance &\multicolumn{1}{c}{Bootstrap}&Pivotal  &\hspace{0.3cm}& \multicolumn{4}{c}{$(n_0,n_1)=(50,20)$}&\hspace{0.2cm}&\multicolumn{4}{c}{$(n_0,n_1)=(200,80)$}\\
  Estimator & Version & Statistic&&$\varrho_{01}=1$&$\varrho_{01}=0$&$\varrho_{01}=-1$&minimum&&
  $\varrho_{01}=1$&$\varrho_{01}=0$&$\varrho_{01}=-1$&minimum\\[4pt]\cline{5-8}\cline{10-13}\\

   $\hmmv_\neyman$ & N/A& No              &&     0.9560  &  0.9656 &   0.9832 &   0.9560  &&  0.9650  &  0.9796  &  0.9880  &  0.9650 \\
    $\hmmv_\aronow$ &N/A&No            &&     0.9352  &  0.9510 &   0.9730 &   0.9352  &&  0.9462  &  0.9664  &  0.9818  &  0.9462 \\
   $\hmmv_\sboot$ &Standard &  No             &&     0.9508  &  0.9616 &   0.9804 &   0.9508  &&  0.9636  &  0.9778  &  0.9878  &  0.9636 \\
   $\hmmv_\cboot$&Causal&  No           &&     0.9308  &  0.9490 &   0.9706 &   0.9308  &&  0.9452  &  0.9654  &  0.9838  &  0.9452 \\[4pt]
   \multicolumn{3}{c}{Fisher's Exact Test}        &&     0.9332  &  0.9112 &   0.8948 &   0.8948  &&  0.8624  &  0.8638  &  0.8616  &  0.8616 \\[4pt]
   $\hmmv_\neyman$ &Standard& Yes            &&     0.9652  &  0.9754 &   0.9878 &   0.9652  &&  0.9660  &  0.9792  &  0.9886  &  0.9660 \\
    $\hmmv_\aronow$ &Standard&Yes         &&     0.9632  &  0.9744 &   0.9878 &   0.9632  &&  0.9656  &  0.9786  &  0.9886  &  0.9656 \\
   $\hmmv_\neyman$ &Causal& Yes           &&     0.9444  &  0.9610 &   0.9776 &   0.9444  &&  0.9492  &  0.9684  &  0.9836  &  0.9492 \\
    $\hmmv_\aronow$ &Causal&Yes           &&     0.9432  &  0.9608 &   0.9774 &   0.9432  &&  0.9490  &  0.9684  &  0.9832  &  0.9490 \\
&\vspace{-0.2cm}\\\cline{5-8}\cline{10-13}\cline{5-8}\cline{10-13}\\
\end{tabular}}\end{center}
\end{table}

In Table \ref{tab:coupling_sim} we report simulated coverage rates for nominal $95\%$ confidence intervals for the average treatment effect, where for either sample size we report the lowest coverage rate across the three different couplings in a separate column. The simulation results broadly confirm the theoretical predictions. Overcoverage from using sampling-based, rather than causal estimators for the variance or the bootstrap is not evident from the design with smaller sample sizes ($n_0=50,n_1=20$), but becomes clearly visible once we move to the design with a larger number of units ($n_0=200,n_1=80)$. The confidence interval based on Fisher's exact test has coverage that is consistently below the nominal $95\%$ level.

Next we compare coverage rates of these confidence intervals as the size of the sample increases, where we choose a design with non-Gaussian distributions for the potential outcomes. Specifically, we let $Y_i(0)\equiv0$ and $Y_i(1)|S_{i}(1)\sim N(0,S_i^2)$, where $S_i = 1$ with probability $0.9$, and $S_i=4$ with probability  $0.1$. Since the marginal distributions for $Y_i(0)$ and $Y_i(1)$ are different, the difference between sampling variance and the upper bound for the causal variance is nontrivial. Furthermore, while we do not give formal results, under certain regularity conditions the pivotal causal bootstrap should be expected to provide refinements over the Gaussian limiting approximation to the randomization distribution.

\begin{table}[ht]\label{tab:refinements_sim}
\caption{Coverage of nominal 95\% Confidence Intervals, non-Gaussian Potential Values with Isotone Coupling}
\begin{center}
\scriptsize{\begin{tabular}{ccccccccccccccc}

Variance &\multicolumn{1}{c}{Bootstrap}&Pivotal  &\hspace{0.3cm}&$(n_0,n_1)$&$(n_0,n_1)$&$(n_0,n_1)$&$(n_0,n_1)$&$(n_0,n_1)$\\
Estimator &Version& Statistic&&   $(20,20)$   & $(50,50)$    & $(100,100)$    & $(200, 200)$   & $(500, 500)$\\[4pt]\cline{5-9}\\

 $\hmmv_\neyman$ &N/A& No              &&   0.9768     &    0.9866    &    0.9914    &    0.9932    &    0.9924 \\
  $\hmmv_\aronow$ &N/A&No            &&   0.9186       &    0.9358    &    0.9396    &    0.9450    &    0.9436 \\
 $\hmmv_\sboot$ &Standard&No             &&   0.9752   &    0.9864    &    0.9912    &    0.9928    &    0.9924 \\
 $\hmmv_\cboot$ &Causal&No           &&   0.9144       &    0.9336    &    0.9378    &    0.9436    &    0.9436 \\[4pt]
 \multicolumn{3}{c}{Fisher's Exact Test}&&   0.9752    &    0.9652    &    0.9672    &    0.9560    &    0.9592 \\[4pt]
 $\hmmv_\neyman$ &Standard&Yes            &&   0.9870  &    0.9912    &    0.9940    &    0.9942    &    0.9934 \\
  $\hmmv_\aronow$ &Standard&Yes         &&   0.9870    &    0.9912    &    0.9940    &    0.9942    &    0.9934 \\
 $\hmmv_\neyman$ &Causal& Yes           &&   0.9470    &    0.9532    &    0.9582    &    0.9548    &    0.9482 \\
  $\hmmv_\aronow$ &Causal&Yes           &&   0.9470    &    0.9532    &    0.9582    &    0.9548    &    0.9482 \\
&\vspace{-0.2cm}\\\cline{5-9}\cline{5-9}\\
\end{tabular}}\end{center}\end{table}

Table \ref{tab:refinements_sim} shows simulated coverage rates for the different confidence intervals at the nominal $95\%$ significance level under this design. The results show that coverage rates for both the sampling-based variance estimators and bootstrap are higher throughout than for their causal analogs. The comparison between Gaussian confidence intervals using the causal variance estimators, $\hmmv_\aronow$ and $\hmmv_\cboot$, respectively, to the pivotal causal bootstrap is also indicative of refinements, where the confidence interval based on the pivotal causal bootstrap has coverage rates much closer to the nominal level for small sample sizes, but that advantage vanishes as $n_0,n_1$ grow large.

\section{Large Sample Theory}

\label{sec:asy_theory_sec}

To characterize the asymptotic properties of the bootstrap procedure, we can cast the statistical experiment of sampling from a finite population with subsequent randomization of treatment among the sampled units as a two-stage scheme of sampling without replacement from nested finite populations. Specifically, in a first step we draw $n$ units without replacement from the population of $N$ units. In a second step, we draw $n_1$ units at random and without replacement from that sample to receive the treatment $W_i=1$, whereas the remaining $n_0=n-n_1$ units are assigned $W_i=0$. This second step is conditionally independent of the first. %In the following we denote $q:=\frac{n}N\equiv\frac1N\sum_{i=1}^NR_i$ and $p=\frac{n_1}n\equiv\frac1n\sum_{i=1}^N R_iW_i$, and for simplicity for now we only describe the case of completely randomized assignment, $Y_i(0),Y_i(1)\ind W_i$.

%\flushleft\textbf{[Q:SHOULD THE FOLLOWING DISCUSSION BE IN TERMS OF THE CDF OR AVERAGES OF $Y_i(0),Y_i(1)$?]}\\[4pt]
To characterize the contribution of sampling uncertainty to the distribution of the functional we define
\begin{eqnarray}
\nonumber F_{01}^p(y_0,y_1)&:=&\frac1N\sum_{i=1}^N\dum\{Y_i(0)\leq y_0,Y_i(1)\leq y_1\}\\
\nonumber F_{01}^s(y_0,y_1)&:=&\frac1n\sum_{i=1}^NR_i\dum\{Y_i(0)\leq y_0,Y_i(1)\leq y_1\}
\end{eqnarray}
with corresponding marginals $F_0^p,F_1^p,F_0^s,F_1^s$. In particular,
\begin{equation}
\label{samp_cdf_diff} F_{01}^s(y_0,y_1)-F_{01}^p(y_0,y_1)=\frac1n\sum_{i=1}^N\left(R_i-q\right)\dum\{Y_i(0)\leq y_0,Y_i(1)\leq y_1\}
\end{equation}

Turning to the contribution of design uncertainty, we define
\begin{eqnarray}
\nonumber\hat{F}_{0}(y_0)&:=&\frac1{n(1-p)}\sum_{i=1}^nR_i(1-W_i)\dum\{Y_i(0)\leq y_0\}\\
\nonumber\hat{F}_{1}(y_1)&:=&\frac1{np}\sum_{i=1}^nR_iW_i\dum\{Y_i(1)\leq y_1\}
\end{eqnarray}
where we can rewrite
\[\hat{F}_{0}(y_0)=\frac{p}{n(1-p)}\sum_{i=1}^NR_i\left(1-\frac{W_i}p\right)\dum\{Y_i(0)\leq y_0\}\]
Hence, we have
\begin{eqnarray}
\label{rand_cdf_diff}\left(\begin{array}{c}\hat{F}_{0}(y_0)-F_0^s(y_0)\\\hat{F}_{1}(y_1)-F_1^s(y_1)\end{array}\right)
&=&\frac1{np}\sum_{i=1}^NR_i(W_i-p)\left(\begin{array}{c} -\frac{p}{1-p}\dum\{Y_i(0)\leq y_0\}\\ \dum\{Y_i(1)\leq y_1\}
\end{array}\right)
\end{eqnarray}

Taken together, (\ref{samp_cdf_diff}) and (\ref{rand_cdf_diff}) characterize the uncertainty from sampling and randomization in estimating the respective marginal distributions of $Y_i(0)$ and $Y_i(1)$ as a two-stage process of drawing without replacement from nested finite populations. An asymptotic Donsker Theorem for empirical processes based on sampling without replacement from a finite population is available from \cite{Bic69}. %We also rely on results by \cite{BGo01} and \cite{BGo02} to establish refinements via finite-population Edgeworth expansions.

We now state the limiting properties of the bootstrap as $N$ and $n$ grow large. Specifically, we derive the limits of the  randomization and bootstrap distributions. We then show that the latter is an asymptotically conservative estimator of the former for the purposes of forming confidence intervals. %For consistency and

\subsection{Consistency and Randomization CLT}

%[TO BE REPLACED: For now, we will state the following auxiliary high-level assumption on $\tau(F_0,F_1),\sigma(F_0,F_1)$ as functionals of the marginal distributions $F_0,F_1$.
%
%\begin{ass}\label{hadamard_diff_ass}\textbf{(Hadamard-Differentiability)} [TO BE REPLACED WITH PRIMITIVE ASSUPTIONS]
%The functionals $\tau_L(F_0,F_1)$ and $\tau_U(F_0,F_1)$ are Hadamard-differentiable with respect to the distributions $F_0,F_1$ with derivatives $\tau_{F_0,F_1;L}'$ and $\tau_{F_0,F_1;U}'$, respectively, and $\sigma_L(F_0,F_1)$ and $\sigma_U(F_0,F_1)$ are continuous in $F_0,F_1$.
%\end{ass}
%We will replace this condition with sufficient primitive conditions on moments of $Y_i(0),Y_i(1)$ to be included with Assumption \ref{SATE_mom_ass} since results in this paper only concern the case of the SATE and its variance. Hadamard differentiability will then be proven as a Lemma in the Appendix.]

Consistency of the estimated bounds follows from consistency of  $\hat{F}_{0}(y_0)$ and $\hat{F}_{1}(y_1)$ for $F_0(y_0)$ and $F_1(y_1)$, respectively, and the continuous mapping theorem, noting that the conditions in Assumption \ref{SATE_mom_ass} are sufficient for the parameter bounds to be continuous functions of $F_0(y_0)$ and $F_1(y_1)$.

\begin{thm}\label{consistency_thm}\textbf{(Consistency)} Suppose Assumptions \ref{SATE_mom_ass}, \ref{density_ass}, and \ref{unconfoundedness_ass} hold. Then $\hat{\tau}$ and $\hat{\sigma}$ are consistent for $\tau(F_0^p,F_1^p)$ and $\sigma(F_0^p,F_1^p)$, respectively.
\end{thm}

%By the main result of \cite{Ima04}, construction of a valid confidence interval of the form \ref{confidence_interval} only requires consistent estimators of the marginal variances $\sigma_L=\sigma_L(F_0^p,F_1^p)$ and $\sigma_U\equiv\sigma_U(F_0^p,F_1^p)$, the distribution of the studentized estimators of the bounds, and the normalized width of the identified set, $\sqrt{n}(\tau_U-\tau_L)$. In particular, we do not attempt to estimate or bound the covariance parameter $\sigma_{UL}$.

For a randomization CLT for the estimated bounds we first establish a functional CLT for the randomization processes
\begin{eqnarray}
\nonumber \hat{G}_{0}&:=&\sqrt{n}(\hat{F}_{0}-F_0^p)\\
\nonumber \hat{G}_{1}&:=&\sqrt{n}(\hat{F}_{1}-F_1^p)
\end{eqnarray}
for conditional distributions of potential outcomes. We argue that Assumption \ref{SATE_mom_ass} is sufficient to establish Hadamard differentiability of the functionals $\tau(\hat{F}_{0},\hat{F}_{1})$, $\sigma(\hat{F}_{0},\hat{F}_{1})$ so that asymptotic normality of $\sqrt{n}\frac{\hat{\tau}-\tau}{\hat{\sigma}}$ follows from the functional Delta rule and Slutsky's theorem.

%A functional CLT for $(\mathbb{G}_{0n},\mathbb{G}_{1n})$ can be obtained by adapting arguments for smoothed empirical processes (see e.g. \cite{Yuk92} and \cite{Vdv94}) to a multiplier CLT that is conditional on the empirical distributions $F_{01}^p(y_0,y_1|x)$ and $F_X^p(x)$. A CLT for the parameters then follows from the Delta rule.

\begin{thm}\label{randomization_clt_thm}\textbf{(Randomization CLT)} Suppose Assumptions \ref{SATE_mom_ass}, \ref{density_ass}, and \ref{unconfoundedness_ass} hold. Then the asymptotic distribution of the t-ratio for $\hat{\tau}_{ATE}$ is given by
\[\sqrt{n}\frac{\hat{\tau}-\tau}{\hat{\sigma}}\stackrel{d}{\rightarrow}N\left(0,\frac{\sigma^2(F_{01}^p)}{\sigma^2(F_0^p,F_1^p)}\right)\]
where $\sigma(F_{01})^2:=\lim_n n\var_{F_{01}}(\hat{\tau})$.
\end{thm}

The proof of this result is given in the appendix. The formal argument adapts a finite-population CLT for the empirical process developed by \cite{Bic69} for the two-sample problem to the case of sampling and randomization in a finite population.

\subsection{Bootstrap CLT}

For a bootstrap replication, denote the empirical distributions of $Y_i^*|W_i^*=0$ and $Y_i^*|W_i^*=1$ with $\hat{F}_{0}^*$ and $\hat{F}_{1}^*$, respectively. Also, let $\hat{\tau}^*=\tau(\hat{F}_{0}^*,\hat{F}_{1}^*)$ and $\hat{\sigma}^*=\sigma(\hat{F}_{0}^*,\hat{F}_{1}^*)$

We then establish a CLT for the bootstrap analogs
\begin{eqnarray}
\nonumber \hat{G}_{0}^*&:=&\sqrt{n}(\hat{F}_{0}^*-\hat{F}_{0})\\
\nonumber \hat{G}_{1}^*&:=&\sqrt{n}(\hat{F}_{1}^*-\hat{F}_{1}).
\end{eqnarray}
A CLT for the bootstrapped bounds $\sqrt{n}\frac{\hat{\tau}^*-\hat{\tau}}{\hat{\sigma}^*}$ then relies again on Hadamard differentiability of the variance bounds and the Delta rule for the bootstrap.

A bootstrap CLT can be shown using analogous steps as in a proof for Theorem \ref{randomization_clt_thm}, where the randomization distribution is generated based on an estimator for $F_{01}^p$ based on the estimated distributions $\hat{F}_0^p(y_0)$, $\hat{F}_1^p(y_1)$ and the respective least-favorable coupling $C^{iso}(\cdot)$.

\begin{thm}\label{bootstrap_clt_thm}\textbf{(Bootstrap CLT)} Suppose Assumptions \ref{SATE_mom_ass}, \ref{density_ass}, and \ref{unconfoundedness_ass} hold. Then the asymptotic distribution of the bootstrapped t-ratio for $\hat{\tau}_{ATE}$ is given by
\[\sqrt{n}\frac{\hat{\tau}^*-\hat{\tau}}{\hat{\sigma}^*}\stackrel{d}{\rightarrow}N\left(0,1\right)\]
\end{thm}

Most importantly, by Theorem \ref{consistency_thm} the bootstrap estimator for the randomization distribution for $\frac{\hat{\tau}-\tau}{\hat{\sigma}}$ has asymptotic variance equal to 1, whereas the asymptotic variance of the randomization distribution is $\frac{\sigma^2(F_{01}^p)}{\sigma^2(F_0^p,F_1^p)}$ which is less than 1 by construction. That is, the bootstrap algorithm in section \ref{sec:algorithm_sec} converges to a ``least-favorable" limiting experiment in an appropriate sense. Note also that the formal argument in the proofs of Theorems \ref{consistency_thm}-\ref{bootstrap_clt_thm} immediately apply to any other functional $\tau(F_0,F_1)$ that is Hadamard-differentiable in $F_0,F_1$, and for which the variance bound $\sigma^2(F_0,F_1)$ is continuous in $F_0,F_1$.

\subsection{Asymptotic Validity of Confidence Intervals}

It remains to show that confidence intervals of the form (\ref{confidence_interval}) that are constructed under the ``least-favorable" limiting experiment are indeed conservative given the CLT under the true randomization distribution in Theorem \ref{randomization_clt_thm}. This can be proven by combining the randomization and bootstrap CLTs, replacing the unidentified randomization variance with an estimate of the bound $\sigma(F_0,F_1)\geq \sigma(F_{01})$.%Also, for the cases of interest for this paper [give sufficient conditions in future draft], the limit of $\sqrt{n}(\tau_U-\tau_L)$ can be taken to be known or consistently estimable.

\begin{cor}\label{conf_validity_thm}\textbf{(Asymptotic Validity of Confidence Intervals)} Under Assumptions \ref{SATE_mom_ass}, \ref{density_ass}, and \ref{unconfoundedness_ass}, the $1-\alpha$ confidence interval (\ref{confidence_interval}) using bootstrap critical values is asymptotically valid,
\[\lim_n \inf_{F_{01}^p}\mathbb{P}_{F_{01}^p}\left(\tau(F_{01}^p)\in CI(1-\alpha)\right)\geq 1-\alpha\hspace{0.5cm}\textnormal{a.s.}\]
\end{cor}

Given Theorems \ref{randomization_clt_thm} and \ref{bootstrap_clt_thm} this result follows immediately from the definition of the variance bound $\sigma(F_0,F_1)$.

\appendix
\footnotesize

\section{Randomization Distribution for $\hat{F}_{0}(y_0),\hat{F}_{1}(y_1)$}

% first unconditional case:

We first compute the randomization covariance $\cov_{W,R}^p(\hat{F}_{0}(y_0),\hat{F}_{1}(y_1))$ given the population distribution $F_{01}^p(y_0,y_1)$, where
\begin{eqnarray}
\nonumber \hat{F}_{0}(y_0)&=&\frac1{n(1-p)}\sum_{i=1}^NR_i(1-W_i)\dum\{Y_i(0)\leq y_0\}\\
\nonumber \hat{F}_{1}(y_1)&=&\frac1{np}\sum_{i=1}^NR_iW_i\dum\{Y_i(1)\leq y_1\}
\end{eqnarray}
In the following we write $A_{0i}:=\dum\{Y_i(0)\leq y_0\}$ and $A_{1i}:=\dum\{Y_i(1)\leq y_1\}$, and take any moments to be with respect to the distribution of $R_i$ and $W_i$ and conditional on the values of $\left(Y_i(0),Y_i(1)\right)_{i=1}^N$ in the population. We then have
\begin{eqnarray}\nonumber\cov(\hat{F}_{0}(y_0),\hat{F}_{1}(y_1))&=&
\frac{1}{n^2p(1-p)}\left[\sum_{i=1}^N\sum_{j=1}^NR_iR_j(1-W_i)W_jA_{0i}A_{1j}\right]\\
\nonumber&=&\frac{1}{n^2p(1-p)}\sum_{i=1}^N\sum_{j=1}^N\mathbb{E}\left[R_iR_j(1-W_i)W_j\right]A_{0i}A_{1j}\\
\nonumber&=&\frac{1}{n^2p(1-p)}\sum_{i=1}^N\sum_{j\neq i}\mathbb{E}\left[R_iR_j\right]\mathbb{E}\left[(1-W_i)W_j\right]A_{0i}A_{1j}\\
\nonumber&=&\frac1{n^2p(1-p)}\sum_{i=1}^N\sum_{j\neq i}\frac{n(n-1)}{N^2}\mathbb{E}\left[(1-W_i)W_j\right]\frac{n^2p(1-p)}{n(n-1)}A_{0i}A_{1j}\\
\nonumber&=&\frac1{N^2}\sum_{i=1}^N\sum_{j\neq i}A_{0i}A_{1j}=\frac1{N^2}\left(\left[\sum_{i=1}^NA_{0i}\right]\left[\sum_{j=1}^NA_{1j}\right]-\sum_{i=1}^NA_{0i}A_{1i}\right)\\
\nonumber&=&-\frac1N\left(F_{01}^p(y_0,y_1)-F_0^p(y_0)F_1^p(y_1)\right)
\end{eqnarray}
To evaluate $\cov\left(\hat{F}_{0}(y_0),\hat{F}_{0}(y_1)\right)$, let $B_{0i}:=\dum\{Y_i(0)\leq y_0\}-F_0^p(y_0)$ and $B_{0i}:=\dum\{Y_i(0)\leq y_1\}-F_0^p(y_1)$. We can then write
\begin{eqnarray}
\nonumber  \cov\left(\hat{F}_{0}(y_0),\hat{F}_{0}(y_1)\right)&=&\frac1{n^2(1-p)^2}\sum_{i=1}^N\sum_{j=1}^N
\mathbb{E}\left[R_iR_j(1-W_i)(1-W_j)\right]B_{0i}B_{1j}\\
\nonumber&=&\frac1{n^2(1-p)^2}\left[\sum_{i=1}^n\frac{n}N\frac{n(1-p)}nB_{0i}B_{1i} + \sum_{i=1}^N\sum_{j\neq i}
\frac{n(n-1)}{N^2}\frac{n(1-p)(n(1-p)-1)}{n^2}B_{0i}B_{1j}\right]\\
\nonumber&=&\frac1{n(1-p)N}\sum_{i=1}^NB_{0i}B_{1i}+\frac{(n-1)(n(1-p)-1)}{n^2(1-p)}\left[\frac1{N^2}\sum_{i=1}^N
\sum_{j\neq i}B_{0i}B_{1j}\right]\\
\nonumber&=&\left[\frac1N\sum_{i=1}^NB_{0i}B_{1i}\right]\left(\frac1{n(1-p)}
-\frac{(n-1)(n(1-p)-1)}{Nn^2(1-p)}\right)\\
\nonumber&=&\left(\min\{F_0^p(y_0),F_0^p(y_1)\} - F_0^p(y_0)F_0^p(y_1)\right)\left(\frac1{n(1-p)}-\frac1N+O\left(\frac1{nN}\right)\right)
\end{eqnarray}
Similarly,
\[\cov\left(\hat{F}_{1}(y_0),\hat{F}_{1}(y_1)\right)=
\left(\min\{F_1^p(y_0),F_1^p(y_1)\} - F_1^p(y_0)F_1^p(y_1)\right)\left(\frac1{np}-\frac1N+O\left(\frac1{nN}\right)\right)
\]

Furthermore,
\begin{eqnarray}
\nonumber  \cov\left(\hat{F}_{0}(y_0),\hat{F}_{0}(y_1)\right)&=&\frac1n\min\{F_0^p(y_0),F_0^p(y_1)\} - F_0^p(y_0)F_0^p(y_1)\\
\nonumber  \cov\left(\hat{F}_{1}(y_0),\hat{F}_{1}(y_1)\right)&=&\frac1n\min\{F_1^p(y_0),F_1^p(y_1)\} - F_1^p(y_0)F_1^p(y_1)
\end{eqnarray}
We let $\mathbf{H}$ denote the covariance kernel of the randomization process with elements
\begin{eqnarray}
\nonumber H_{00}(y_0,y_0')&=&\lim_n n\cov(\hat{F}_{0}(y_0),\hat{F}_{0}(y_0')) = \left(\frac1{1-p}-\frac{n}N\right)\left(\min\{F_0^p(y_0),F_0^p(y_0')\}-F_0^p(y_0)F_0^p(y_0')\right)\\
\label{rand_cov_kernel} H_{01}(y_0,y_1)&=&\lim_n n\cov(\hat{F}_{0}(y_0),\hat{F}_{1}(y_1')) =
\lim_n\frac{n}N\left(F_{01}^p(y_0,y_1) - F_0^p(y_0)F_1^p(y_1)\right)\\
\nonumber H_{11}(y_1,y_1')&=&\lim_n n\cov(\hat{F}_{1}(y_1),\hat{F}_{1}(y_1')) = \left(\frac1{p}-\frac{n}N\right)\left(\min\{F_1^p(y_1),F_1^p(y_1')\}-F_1^p(y_1)F_0^p(y_1')\right)
\end{eqnarray}
Note also that $\frac1{1-p}\geq1\geq\frac{n}N\geq0$ and $\frac1p\geq1\geq\frac{n}N\geq0$, so that $H_{00}(\cdot,\cdot)$ and $H_{11}(\cdot,\cdot)$ are nonnegative.

\section{Proofs for Section \ref{sec:SATE_section}}

\label{sec:SATE_proofs_app_sec}

\subsection{Least Favorable Coupling for the Average Treatment Effect}

%We find that the isotone coupling $Y_i(0)^{iso},Y_i(1)^{iso}$ results in a conservative estimate of the distribution in an appropriate sense. Specifically, we order the distributions for the estimated ATE according to second-order stochastic dominance:
%
%\begin{dfn}\textbf{(Mean-Preserving Spread)} For a pair of random variables $W,W'$, $W'$ is called a \emph{mean-preserving spread} of $W$ if there exists a random variable $Z$ with $\mathbb{E}[Z|W]=0$ a.s. and $W+Z$ has the same distribution as $W'$. We also say that $W'$ \emph{stochastically dominates} $W$ to \emph{second order}, in symbols $W'\succsim_{SOSD}W$.
%\end{dfn}
%
%It is then straightforward to show that the isotone coupling of potential outcomes results in a distribution for the $\sqrt{n}(\hat{\tau}_{ATE}-\tau_{SATE})$ that stochastically dominates the randomization distribution under any other assumption on the copula for the potential outcome distribution.

We first prove a more general result than Proposition \ref{sate_iso_prp} by showing that the isotone coupling of potential outcomes in fact results in a distribution for the ATE parameter which dominates that under any other coupling in the sense of second-order stochastic dominance (SOSD):

\begin{lem}\label{ate_sosd_lem}\textbf{(Ordering of Distributions)} %Suppose $Y_0,Y_0^{iso}\sim F_0$ and $Y_1,Y_1^{iso}\sim F_1$, where the joint distribution of $(Y_0^{iso},Y_1^{iso})$ has the copula $C(u,v) = \min\{u,v\}$. Also, let
%\begin{eqnarray}\nonumber\hat{\tau}_{ATE}&:=&\frac1{np}\sum_{i=1}^nW_iY_{1i} - \frac1{n(1-p)}\sum_{i=1}^n(1-W_i)Y_{0i}\\
%\nonumber\hat{\tau}_{ATE}^{iso}&=&\frac1{np}\sum_{i=1}^nW_iY_{1i}^{iso} - \frac1{n(1-p)}\sum_{i=1}^n(1-W_i)Y_{0i}^{iso}
%\end{eqnarray}
Let $F_{01}$ be an arbitrary joint distribution with marginal distributions $F_0$ and $F_1$, and let $F_{01}^{iso}:=C^{iso}(F_0,F_1)$ be the joint distribution under the isotone coupling. Then for any convex function, the randomization distribution for $\hat{\tau}_{ATE}$ satisfies
\[\mathbb{E}_{F_{01}^{iso}}[v(\hat{\tau}_{ATE})] \geq \mathbb{E}_{F_{01}}[v(\hat{\tau}_{ATE})]\]
For any strictly convex function $v(\cdot)$ this inequality is strict whenever $F_{01}\neq F_{01}^{iso}$.
%Then for any distribution of $W_1,\dots,W_n$, we have $\hat{\tau}_{ATE}^{iso}\succsim_{SOSD}\hat{\tau}_{ATE}$.
\end{lem}

This result is a straightforward consequence of the familiar observation that the isotone (assortative) coupling of potential outcomes results in the distribution of $Y_i(0)+Y_i(1)$ which second-order stochastic dominates that resulting from any other coupling (see e.g. \cite{Bec73}, \cite{FPa10}, and \cite{Sto10}). For illustrative purposes, we give a complete proof here.

\textsc{Proof:} In order to establish second-order stochastic dominance of the isotone assignment $Y_i(1)=F_1^{-1}(F_0(Y_i(0)))$, consider the expectation of $v(\hat{\tau}_{ATE})$ for any convex function $v(u)$. Note that for any pair of observations $i,j$ we can write
\[\hat{\tau}_{ATE} = \frac1n\left(B_{-ij} + R_iW_i\left(Y_i(0)/(1-p) + Y_i(1)/p\right) +
R_jW_j\left(Y_j(0)/(1-p) + Y_j(1)/p\right)\right) \]
where $B_{-ij}:=\sum_{k\neq i,j}R_k\left(W_k(Y_k(0)/(1-p) + Y_k(1)/p) - Y_k(0)/(1-p)\right)  -
(Y_i(0) + Y_j(0))/(1-p)$.

We can now consider the change in $\mathbb{E}[v(\hat{\tau}_{ATE})]$ from
pairwise substitutions of potential outcomes between units $i$ and $j$. Specifically suppose that under the initial coupling, the potential outcomes for unit $i$ are given by $Y_i(0),Y_i(1)$, and the potential outcomes for unit $j$ are $Y_j(0),Y_j(1)$. We then consider the effect of switching the assignment to potential outcomes $Y_i(0),Y_j(1)$ for unit $i$, and potential outcomes $Y_j(0),Y_i(1)$ for unit $j$.

Since $W_i,W_j$ are independent of $W_k$, that change leads to an increase in $\mathbb{E}[v(\hat{\tau}_{ATE})]$ if and only if
\begin{eqnarray}0&\leq& \nonumber P(W_i=1,W_j=0)\left\{v(B_{-ij}+Y_{i}(0)/(1-p)+Y_{i}(1)/p)-v(B_{-ij}+Y_{i}(0)/(1-p)+Y_{j}(1)/p)\frac{}{}\right\}\\
\nonumber&&+P(W_i=0,W_j=1)\left\{v(B_{-ij}+Y_{j}(0)/(1-p)+Y_{j}(1)/p)-v(B_{-ij}+Y_{j}(0)/(1-p)+Y_{i}(1)/p)\frac{}{}\right\}\\
\nonumber&=&p(1-p)\left\{\frac{}{}v(B_{-ij}+Y_{i}(0)/(1-p)+Y_{i}(1)/p)+v(B_{-ij}+Y_{j}(0)/(1-p)+Y_{j}(1)/p)\right.\\
\nonumber&&\left.-v(B_{-ij}+Y_{i}(0)/(1-p)+Y_{j}(1)/p)-v(B_{-ij}+Y_{j}(0)/(1-p)+Y_{i}(1)/p)\frac{}{}\right\}\end{eqnarray}
for any pair of observations with $R_i=R_j=1$. Noting that for any convex function $v(\cdot)$, $v(b+x_0+x_1)$ is supermodular in $x=(x_0,x_1)'$, this difference is nonnegative if and only if $Y_i(0)-Y_j(0)$ and $Y_i(1)-Y_j(1)$ have the same sign. Furthermore, if in addition $v(\cdot)$ is strictly convex, the first inequality is strict.

Since any coupling of potential outcomes can be obtained from the isotone assignment by pairwise substitutions of this form, the isotone assignment maximizes the expectation \[\mathbb{E}[v(\hat{\tau}_{ATE})] = \mathbb{E}\left[v\left(\frac1n\sum_{i=1}^NR_i\left\{W_iY_i(1)/p-(1-W_i)Y_i(0)/(1-p)\right\}\right)\right]\] for all convex functions $v(\cdot)$. Therefore the distribution of $\hat{\tau}_{ATE}$ under the isotone assignment dominates that under any alternative coupling, as claimed above. \qed

\subsubsection*{Proof of Proposition \ref{sate_iso_prp}:} The claim in the proposition follows immediately from Lemma \ref{ate_sosd_lem} and the observation that the function $v(y)=y^2$ is  strictly convex\qed

\section{Proofs for Section \ref{sec:asy_theory_sec}}

\label{sec:asy_proofs_app_sec}

\subsubsection{Proof of Theorem \ref{consistency_thm}} From standard results (see e.g. Example 19.6 in \cite{Vdv98}), the class $\mathcal{F}:=\{(-\infty,y]:y\in\mathbb{R}\}$ is Glivenko-Cantelli, so that $(\hat{F}_{0}-F_0^p,\hat{F}_{1}-F_1^p)$ converges to zero almost surely as an element of the space of bounded functions on $\mathbb{R}$. Since Assumption \ref{SATE_mom_ass} is sufficient to guarantee that the functionals $\tau(F_0,F_1)$ and $\sigma(F_0,F_1),$ are continuous in $F_0,F_1$, the claim of the Theorem follows immediately from the continuous mapping theorem (see e.g. Theorem 18.11 in \cite{Vdv98})\qed

%\subsection{Proof of Theorem \ref{randomization_0lt_thm}}
\vspace*{0.4cm}
For the proof of Theorem \ref{randomization_clt_thm}, we need to characterize functional convergence of the randomization process. To that end, we first introduce some standard notation from empirical process theory (see \cite{Vvw96}). Let $\mathcal{F}:=\{\dum\{y\leq (-\infty,t]\}:t\in\mathbb{R}\}$ be the class of indicator functions for the left-open half-lines on $\mathbb{R}$ and let $\ell^{\infty}(\mathcal{F})$ be the space of bounded functions from $\mathcal{F}$ to $\mathbb{R}$ endowed with the norm $\|z\|_{\mathcal{F}}:=\sup_{f\in\mathcal{F}}|z(f)|$. Also, let $BL_1$ denote the set of all functions $h:\ell^{\infty}{\mathcal{F}}\mapsto[0,1]$ with $|h(z_1)-h(z_2)|\leq \|z_1-z_2\|_{\mathcal{F}}$.

%results for known propensity score $p$, later generalize

\begin{lem}\label{rand_mult_clt_lem}
Suppose that $(Y_i(0),Y_i(1))\stackrel{iid}{\sim} F_{01}$. Then the randomization process \[\hat{G}_n:=\sqrt{n}\left(\begin{array}{c}\hat{F}_{0}-F_0^p\\ \hat{F}_{1}-F_1^p\end{array}\right)\]
converges in outer probability to $\mathbb{G}$ under the bounded Lipschitz metric,
\[\sup_{h\in BL_1}|\mathbb{E}_{W}h(\hat{G}_n)-\mathbb{E}h(\mathbb{G})|\rightarrow0\]
in outer probability, where $\mathbb{G}$ is a Gaussian process with covariance kernel $\mathbf{H}$.
\end{lem}

% use formulation from bootstrap CLT?

\textsc{Proof:} As before, denote the joint c.d.f. of potential outcomes (observed and counterfactuals) for the $n$ units included in the sample with
\[F_{01}^s(y_0,y_1):=\frac1n\sum_{i=1}^NR_i\dum\{Y_i(0)\leq y_0,Y_i(1)\leq y_1\}\]
and the empirical c.d.f. among the units included in the sample for which $W_i=1$,
\[F_{01}^{t}(y_0,y_1):=\frac1{np}\sum_{i=1}^NR_iW_i\dum\{Y_i(0)\leq y_0,Y_i(1)\leq y_1\}\]
Using this notation we can write
\[\sqrt{n}(F_{01}^t(y_0,y_1)-F_{01}^p(y_0,y_1))=\sqrt{n}(F_{01}^t(y_0,y_1)-F_{01}^s(y_0,y_1))
+\sqrt{n}(F_{01}^s(y_0,y_1)-F_{01}^p(y_0,y_1))\]

Since $R_i,W_i$ are drawn at random and without replacement, it follows from Theorem 3.1 in \cite{Bic69} that
\begin{eqnarray}
\nonumber\sqrt{n}(F_{01}^t(y_0,y_1)-F_{01}^s(y_0,y_1))&\rightsquigarrow&\mathbb{G}_{F_{01}^s}\\
\nonumber\sqrt{n}(F_{01}^s(y_0,y_1)-F_{01}^p(y_0,y_1))&\rightsquigarrow&\mathbb{G}_{F_{01}^p}
\end{eqnarray}
for Brownian bridges $\mathbb{G}_{F_{01}^s}$ and $\mathbb{G}_{F_{01}^p}$. Since for any joint distribution $F_{01}(y_0,y_1)$ the marginals satisfy $\lim_{y_1\rightarrow\infty}F_{01}(y_0,y_1)=F_0(y_0)$ for each $y_0$, weak convergence of the joint process implies weak convergence of the marginal empirical processes,
\begin{eqnarray}
\nonumber\sqrt{n}(F_{0}^t-F_{0}^p)&\rightsquigarrow&\mathbb{G}_{F_{0}^s}+\mathbb{G}_{F_{0}^p}\\
\nonumber\sqrt{n}(F_{1}^t-F_{1}^p)&\rightsquigarrow&\mathbb{G}_{F_{1}^s}+\mathbb{G}_{F_{1}^p}
\end{eqnarray}
Finally, $\hat{F}_{1}(y_1)\equiv F_{1}^t(y_1)$ and $\hat{F}_{0}(y_0)\equiv \frac{1}{(p-1)}(F_{0}^s(y_0) - pF_{0}^t(y_0))$, establishing the claim, where the structure of the covariance kernel follows from the point-wise calculations in the derivation of (\ref{rand_cov_kernel}) \qed

% marginal c.d.f.s

%We can write the randomization process as
%\begin{eqnarray}
%\nonumber\hat{F}_{0}(y_0) - F_0^p(y_0) &=& \frac1{n(1-p)}\sum_{i=1}^n\left(p-W_i\right)\dum\{Y_i(1) - y_1\}\\
%\nonumber\hat{F}_{1}(y_1) - F_1^p(y_1) &=& \frac1{np}\sum_{i=1}^n\left(W_i-p\right)\dum\{Y_i(1) - y_1\}
%\end{eqnarray}
%Now let $Z_i:=(Y_i(0),Y_i(1))'$ and $\xi_i:=W_i-\hat{p}_n$. Since $\mathcal{F}$ is Donsker, it follows from a conditional multiplier CLT (e.g. Theorem 2.9.6 in \cite{Vvw96}) that the process
%\[\sqrt{n}\left(\begin{array}{c}-\frac1{n(1-p)}\sum_{i=1}^n\xi_i(\delta_{Y_i(0)}-F_0)\\
%\frac1{np}\sum_{i=1}^n\xi_i(\delta_{Y_i(1)}-F_1)\end{array}\right)\rightsquigarrow\mathbb{G}\]
%in outer probability with respect to random sequences $Z_1,\dots,Z_n$. Note that weak convergence here is understood to be with respect to the bounded Lipschitz metric over the set of bounded functions $\ell^{\infty}(\mathcal{F})$ \qed

\subsubsection{Proof of Theorem \ref{randomization_clt_thm}:} From Assumption \ref{SATE_mom_ass} it is immediate that $\tau(F_0,F_1)$ is Hadamard-differentiable.  Lemma \ref{rand_mult_clt_lem} and the functional delta method, see e.g. Theorem 20.8 in \cite{Vdv98}, then imply asymptotic normality of $\sqrt{n}(\hat{\tau}-\tau)/\sigma(F_0,F_1)$. Theorem \ref{randomization_clt_thm} then follows from Slutsky's theorem and consistency of $\hat{\sigma}$ from Theorem  \ref{consistency_thm}\qed

\vspace*{0.4cm}

We next turn to the bootstrap distribution: Consider the bootstrap replicates
\begin{eqnarray}
\nonumber \hat{F}_{0}^*(y_0):=\frac1{n(1-p)}\sum_{i=1}^nR_i^*(1-W_i^*)\dum\{Y_i^*(0)\leq y_0\},&\hspace{0.3cm}&\hat{F}_{1}^*(y_1):=\frac1{np}\sum_{i=1}^nR_i^*W_i^*\dum\{Y_i^*(1)\leq y_1\}
\end{eqnarray}
by randomizing from $\hat{F}_{01}$. We also define the asymptotic covariance kernel $\mathbf{H}^{iso}$ corresponding to the coupling $C^{iso}$ in analogy to (\ref{rand_cov_kernel}) where  $F_{01}$ is replaced with $C^{iso}(F_0,F_1)$. We first show the two following Lemmas:

\begin{lem}\label{rand_consistency_lem} Suppose that $(Y_i(0),Y_i(1))\stackrel{iid}{\sim} F_{01}$. Then for any copula $C:[0,1]^2\rightarrow[0,1]$,
\[\sup_{y_0,y_1\in\mathbb{R}}\left|C(\hat{F}_{0},\hat{F}_{1})(y_0,y_1) - C(F_0^p,F_1^p)(y_0,y_1)\right|\stackrel{a.s.}{\rightarrow}0\]
\end{lem}

\textsc{Proof:} From standard results, the class $\mathcal{F}:=\{(-\infty,y]:y\in\mathbb{R}\}$ is Glivenko-Cantelli, so that $(\hat{F}_{0}-F_0^p,\hat{F}_{1}-F_1^p)$ converges to zero almost surely as an element of the space of bounded functions on $\mathbb{R}$. Noting that any copula  $C:[0,1]^2\rightarrow[0,1]$ is a bounded nondecreasing function in each of its arguments, it follows that
\[\sup_{y_0,y_1\in\mathbb{R}}\left|C(\hat{F}_{0},\hat{F}_{1})(y_0,y_1) - C(F_0^p,F_1^p)(y_0,y_1)\right|\stackrel{a.s.}{\rightarrow}0\]
establishing the claim\qed

\begin{lem}\label{bootstrap_clt_lem}
Suppose that $(Y_i(0),Y_i(1))\stackrel{iid}{\sim} F_{01}$. Then the bootstrap process
\[\hat{G}_{n}^*:=\sqrt{n}\left(\begin{array}{c}\hat{F}_{0}^*-\hat{F}_{0}\\\hat{F}_{1}^*-\hat{F}_{1}\end{array}\right)\]
converges in outer probability to $\mathbb{G}$ under the bounded Lipschitz metric, that is
\begin{eqnarray}
\nonumber \sup_{h\in BL_1}\left|\mathbb{E}_{W}h(\hat{G}_{n}^*)-\mathbb{E}h(\mathbb{G})\right|&\rightarrow&0
\end{eqnarray}
in outer probability, where $\mathbb{G}$ is a Gaussian processes with covariance kernel $\mathbf{H}$.
\end{lem}

\textsc{Proof:} By construction of the coupling $(Y_i^*(0),Y_i^*(1))$, the marginal distributions of $Y_i^*(0)$ and $Y_i^*(1)$ are equal to $\hat{F}_{0}$ and $\hat{F}_{1}$, respectively. By construction of the bootstrap, the bootstrap replications $\hat{F}_{0}^*,\hat{F}_{1}^*$ are generated by randomization from the samples $(\tilde{Y}_{i}(1),\tilde{Y}_i(1))_{i=1}^n$
corresponding to the joint distribution $\hat{F}_{01}:=C^{iso}(\hat{F}_{0},\hat{F}_{1})$.

Now let $\hat{\mathbf{H}}^{iso}$ the covariance kernel obtained from (\ref{rand_cov_kernel}) replacing $F_0$ with $\hat{F}_{0}$, $F_1$ with $\hat{F}_{1}$, and $F_{01}$ with $C^{iso}(\hat{F}_{0},\hat{F}_{1})$, respectively. By construction, the bootstrap distribution of $\hat{G}_{n}^*$ conditional on $\hat{F}_{0},\hat{F}_{1}$ have covariance given by $\hat{\mathbf{H}}^{iso}$. Finally, $\hat{\mathbf{H}}^{iso}$ is a continuous function of $C^{iso}(\hat{F}_{0},\hat{F}_{1})$. Hence by Lemma \ref{rand_consistency_lem} and the continuous mapping theorem we have that \[\|\hat{\mathbf{H}}^{iso}-\mathbf{H}^{iso}\|\stackrel{a.s.}{\rightarrow}0\]
which completes the proof.

The claim of the Lemma then follows from the same arguments as in Lemma \ref{rand_mult_clt_lem} and the continuous mapping theorem\qed

\subsubsection{Proof of Theorem \ref{bootstrap_clt_thm}:} It follows from Assumption \ref{SATE_mom_ass} that $\tau(F_0,F_1),\sigma(F_0,F_1)$ are Hadamard differentiable, so that Theorem \ref{bootstrap_clt_lem} follows from Lemma \ref{bootstrap_clt_lem} and the functional Delta method (e.g. Theorem 20.8 in \cite{Vdv98})\qed

%\subsection{Proof of Corollary \ref{conf_validity_thm}}

% Given Theorems \ref{randomization_0lt_thm} and \ref{bootstrap_clt_thm} and the definition of the variance bounds $\sigma_L(F_0,F_1)$ and $\sigma_U(F_0,F_1)$, this result follows immediately from Lemma 4 in \cite{Ima04}.

% results from Imbens and Manski Lemma 4

\normalsize

\bibliographystyle{econometrica}
\bibliography{mybibnew}

\end{document}